\documentclass{ws-ijmpc}

\usepackage{graphicx}
\usepackage{epstopdf}
\usepackage{amsmath}
\usepackage{braket}
\usepackage{siunitx}

\newcommand{\ie}{\emph{i.e.}}
\newcommand{\eg}{\emph{e.g.}}
\newcommand{\etal}{\emph{et al.}}
\newcommand{\rmd}{\mathrm{d}}
\newcommand{\rme}{\mathrm{e}}
\newcommand{\rmi}{\mathrm{i}}

\begin{document}

\title{Wave-packet dynamics of an atomic ion in a Paul trap:
  approximations and stability}

\author{A. Hashemloo$^*$, C. M. Dion$^{*,\dag}$ and G. Rahali$^{*,\ddag}$}
\address{$^*$Department of Physics \\ Ume{\aa} University, SE-901\,87 Ume{\aa},
  Sweden \\
$^{\ddag}$Department of Physics \\ Jazan University, Jazan, Kingdom
 of Saudi Arabia \\
$^\dag$claude.dion@physics.umu.se
}

\maketitle

\begin{abstract}
  Using numerical simulations of the time-dependent Schr\"{o}dinger
  equation, we study the full quantum dynamics of the motion of an
  atomic ion in a linear Paul trap.  Such a trap is based on a
  time-varying, periodic electric field, and hence corresponds to a
  time-dependent potential for the ion, which we model exactly.  We
  compare the center of mass motion with that obtained from classical
  equations of motion, as well as to results based on a
  time-independent effective potential.  We also study the
  oscillations of the width of the ion's wave packet, including close
  to the border between stable (bounded) and unstable (unbounded)
  trajectories.  Our results confirm that the center-of-mass motion
  always follow the classical trajectory, that the width of the wave
  packet is bounded for trapping within the stability region, and
  therefore that the classical trapping criterion are fully applicable
  to quantum motion.

\end{abstract}

\ccode{PACS Nos.: 37.10.Ty,	
02.60.Cb 
}


\section{Introduction}

A Paul trap consists of a series of electrodes creating a
time-dependent (radio-frequency) electric field that enables the
trapping of charged particles (specifically
ions).\cite{Paul_RMP_1990,Ghosh_book_1995,Major_book_2005}  In its
linear configuration,\cite{Raizen_JMO_1992,Raizen_PRA_1992} it
provides the capacity of trapping many ions simultaneously along an
axis.  The linear Paul trap is suitably designed for quantum computing
experiments, reducing the Coulomb repulsion between the ions at the
center of the trapping region, reducing a possible increase in
amplitude of the micromotion of the ion.\cite{Raizen_JMO_1992} Also,
the spacing between electrodes and spacing between ions provide a good
environment for laser controlling experiments.\cite{Brkic_PRA_2006}

The difficulty of a numerical treatment of the quantum dynamics of a
trapped ion stems from the fact that the trapping potential is time
dependent, on a time scale that can be considered slow with respect to
the internal dynamics of the ion, and that the motion of the center of
mass and the width of the wave function can cover greater than
micrometer-sized regions.  Wave packet dynamics of trapped ions
require thus large spatial grids and long simulation times.
Consequently, numerical simulations of the ion motion in the trap have
mostly been limited to classical trajectories (see, \eg,
Refs.~\refcite{Brkic_PRA_2006,Dawson_JVST_1968,Lunney_JAP_1989,%
  Londry_JASMS_1993,Reiser_IJMSIP_1992,Wu_JASMS_2006,Wu_IJMS_2007,%
  Herbane_IJMS_2011,Cetina_PRL_2012}), or have rested on the effective
potential approximation,\cite{Cook_PRA_1985} treating the trapping
potential as purely harmonic (see, \eg,
Refs.~\refcite{Gardiner_PRA_1997,Zheng_PLA_1998,Moya-Cessa_PR_2012}).
There have been only limited studies of the quantum dynamics with the
full time-dependent potential, either looking at the breathing of a
wave packet in the center of the trap,\cite{Li_PRA_1993} or
comparisons between quantum trajectories and an effective potential
approximation.\cite{Major_book_2005} While there exists what can be
called analytical solutions for the center-of-mass motion or the width
of the wave
packet,\cite{Combescure_AIHPA_1986,Brown_PRL_1991,Glauber_1992,Glauber_1992b,Leibfried_RMP_2003,Li_PRA_1993}
simulations based on these formulas must ultimately rely on involved
numerical calculations.  One notable exception is an approximate
solution for a breathing wave packet,\cite{Leibfried_RMP_2003} which
we will address below.

In this paper, we present a full investigation of the
quantum-mechanical motion of an atomic ion in a linear Paul trap,
using the actual time-dependent trapping potential, using numerical
simulations of the spatial wave packet dynamics, focusing our
attention on both the trajectory of the center of mass and the width
of the wave packet.  We compare our results with classical simulations
of the ion motion and examine the validity of the effective potential
approximation.  We also look at the applicability of classical
trapping stability criteria to quantum motion.  Indeed, while a
Floquet analysis\cite{Combescure_AIHPA_1986,Brown_PRL_1991} and the
study of Ref.~\refcite{Li_PRA_1993} point to the applicability of the
classical criteria in all cases, Wang \etal\cite{Wang_PRA_1995} have
claimed that there are trapping parameters for which a quantum
trajectory is stable, while the classical counterpart would not be.

This paper is arranged as follows.  We start by presenting the models,
quantum and classical, for the dynamics of an atomic ion in a linear
Paul trap.  The numerical methods corresponding to both models are
presented in Sec.~\ref{sec:num}.  This is followed by the results of
the various simulations.  Finally, concluding remarks are given in
Sec.~\ref{sec:conclusion}.

\section{\label{sec:model}Model for an atomic ion in a linear Paul
  trap}

\subsection{Hamiltonian of an atomic ion in a linear Paul trap}

We consider a linear Paul trap, such as the one described in
Refs.~\refcite{Berkeland_JAP_1998,Berkeland_PRL_1998,Rohde_JOB_2001},
constructed of four cylindrical electrodes, each located at the corner
of a square. A pair of electrodes that are opposing each other
diagonally is attached to a radio-frequency source and the other pair
is grounded. Also, in order to confine ions inside the trapping
device, a static potential is applied to two ring-shaped electrodes,
located near the end of the cylindrical electrodes.  The
time-dependent electric potential near the center of a linear Paul
trap is given by
\begin{equation} \label{eq:pot}
\Phi(t) = \Phi_{\mathrm{rf}}(t) +\Phi_{\mathrm s},
\end{equation}
where $\Phi_{\mathrm{rf}}(t)$ is a quadrupolar, time-dependent
electric potential of the form
\begin{equation} 
  \Phi_{\mathrm {rf}}(t) = \frac{1}{2} \frac{V_{0}}{r_{0}^{2}}\left (
    x^{2} - y^{2} \right ) \cos \Omega t, 
 \label{eq:rf-pot}
\end{equation}
with $V_{0}$ is the amplitude of the radio-frequency potential of
frequency $\Omega$ and $r_{0}$ is the minimum distance from the
electrodes to the central (trap) axis.

The form of the static potential is dependent on the configuration of
the system, but it is approximated to have a quadratic dependence on
spatial coordinates, especially in the region close to the trapping
axis,\cite{Raizen_PRA_1992,Berkeland_JAP_1998} leading to
\begin{equation} 
  \Phi_{\mathrm s} = \frac{\kappa U_{0}}{z_{0}^{2}}\left [ z^{2} -
    \frac{1}{2}\left ( x^{2} + y^{2} \right ) \right ],
  \label{eq:st-pot}
\end{equation}
in which $U_{0}$ is the amplitude of the static potential, $\kappa$ is
the geometric factor which is a parameter that can be found
experimentally from the oscillation frequency of an ion in the
trap,\cite{Raizen_PRA_1992} and $z_{0}$ is half the distance between
ring-shaped electrodes, along the trapping axis.

Now considering an atomic ion in such an external trapping field, we
have the Hamiltonian for the motion of the ion as
\begin{equation}  
  \hat{H} = -\frac{\hbar^{2}}{2 m} \nabla_{\mathbf{r}}^{2} + Z e \Phi,
  \label{eq:hamiltoni}
\end{equation}
where $m$ is the mass of the ion with charge $Z$, $\mathbf{r}$ its
position vector, $\hbar$ the reduced Planck constant, and $e$ the
elementary charge.

\subsection{Classical trajectories and stability conditions}
\label{sec:stability}

From the Hamiltonian of the system for an ion in a linear Paul trap,
we can find the classical trajectories of the motion of the center of
mass.  Using the standard
approach,\cite{Ghosh_book_1995,Major_book_2005,Drewsen_PRA_2000} we
write the classical equation of motion of an ion in a linear Paul trap
as
\begin{equation} 
  m \ddot{\mathbf{r}} + Ze\vec{\nabla}\Phi = 0
  \label{eq:eq_motion}
\end{equation}
and rewriting it explicitly in its components
\begin{subequations}
  \label{eq:eq_motion_3D}
  \begin{align} 
    m\ddot{x} + Ze \left[ - \frac{\kappa U_{0}}{z_{0}^{2}} +
      \frac{V_{0}}{r_{0}^{2}}\cos \Omega t \right] x &= 0, \\ 
    m\ddot{y} + Ze \left[ - \frac{\kappa U_{0}}{z_{0}^{2}} -
      \frac{V_{0}}{r_{0}^{2}}\cos \Omega t \right] y &= 0, \\ 
    m\ddot{z} + Ze \left[  \frac{2\kappa U_{0}}{z_{0}^{2}} \right] z &= 0, 
  \end{align}
\end{subequations}
we obtain the Mathieu equation
\begin{equation} 
  \frac{\rmd^{2} \mathbf{r} }{\rmd \tau^{2} } + (a - 2q\cos
  2\tau )\mathbf{r} = 0, 
  \label{eq:Math.eq}
\end{equation}
by setting $\tau = \Omega t/2$ and
\begin{subequations}
  \label{eq:a and q}
  \begin{align} 
    a_{x} = a_{y} &= -\frac{4Ze}{m\Omega^{2}}\frac{\kappa U_{0}}{z_{0}^{2}},\\
    a_{z} &= \frac{8Ze}{m\Omega ^{2}}\frac{\kappa U_{0}}{z_{0}^{2}},\\
    q_{x} = -q_{y} &= -\frac{2Ze}{m\Omega^{2}} \frac{V_{0}}{r_{0}^{2}},\\
    q_{z} &= 0\, .
  \end{align}
\end{subequations}

\begin{figure}
  \includegraphics[width=0.49\textwidth]{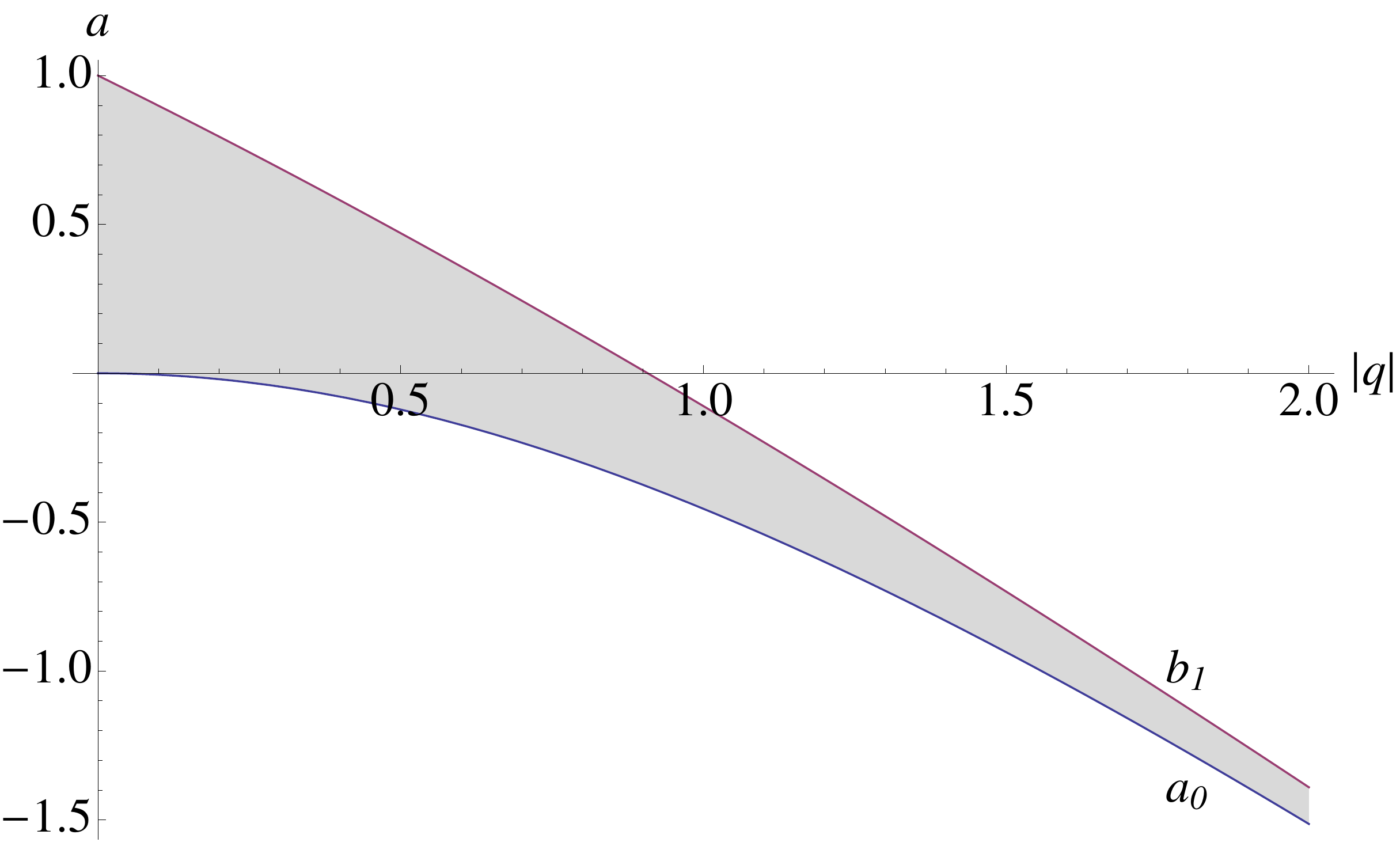}
  \includegraphics[width=0.49\textwidth]{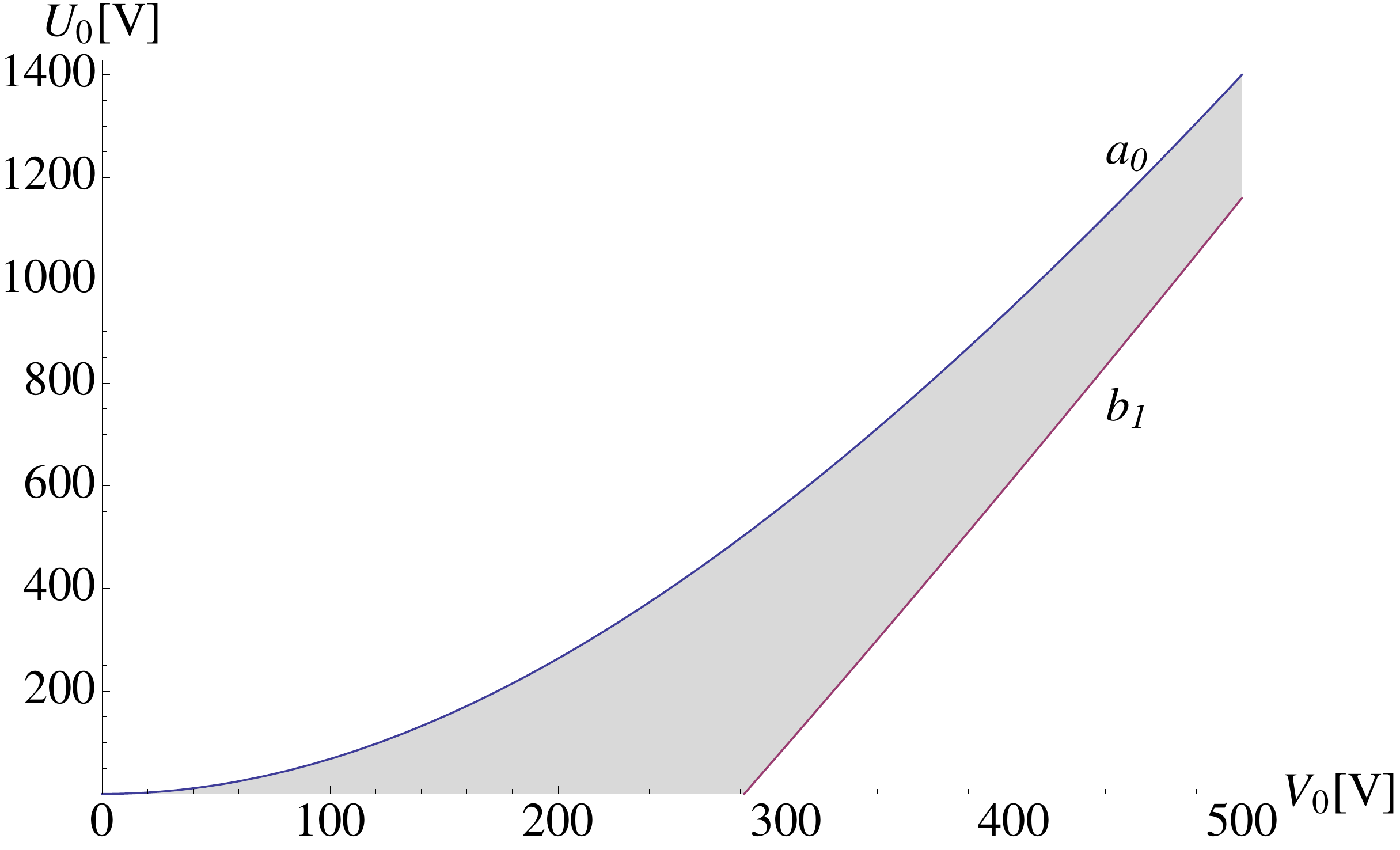}
  \caption{\label{fig:stability}(Color online) Left panel: First
    stability region of the Mathieu equation~(\ref{eq:Math.eq}),
    bounded by the characteristic values $a_0$ and
    $b_1$,\protect\cite{Wolf_NIST_2010} 
    for motion in the xy plane.  Right
    panel: Stability diagram of a linear Paul trap for the full 3D
    motion for a Ca$^+$ ion.  In both panels the shaded region
    corresponds to stable trajectories.}
\end{figure}

Stable solutions of the Mathieu equation, corresponding to a trapped
ion, exist for certain regions in the $a$--$q$
plane.\cite{Wolf_NIST_2010}  We will consider potential values $U_0$
and $V_0$ close to the so-called first stability region, illustrated
in Fig.~\ref{fig:stability}.  

\subsection{Effective potential approximation}
\label{sec:effpot}

The bounded solutions to the ion motion, Sec.~\ref{sec:stability}, are
periodic and consist of two types of motions, an average secular
motion on which a high-frequency micromotion is
superposed.\cite{Ghosh_book_1995,Major_book_2005} The secular motion
corresponds to the trajectory which should be observed in the
time-average electric potential and the small-amplitude micromotion
is driven at the frequency of the oscillation of the potential.  As
the time-dependent trapping field gives rise to approximately harmonic
secular motion of a particle in all
directions,\cite{Leibfried_RMP_2003} specifically near the center of
the trap,\cite{Raizen_PRA_1992} the problem can be approximated using
a harmonic oscillator potential.\cite{Major_book_2005,Cook_PRA_1985}
This approximation is called effective potential approximation or
adiabatic approximation, and is used to calculate an approximated wave
function,\cite{Major_book_2005} which can be compared to the actual
wave function.

The equation of the motion for the ion can be written
as\cite{Major_book_2005}
\begin{equation} 
  m\ddot{\mathbf{r}} = \mathbf{F}_{\mathrm{sm}}(\mathbf{r}) +
  \mathbf{F}_{\mathrm{mm}}(\mathbf{r},t), 
  \label{eq:motion}
\end{equation}
where $\mathbf{F}_{\mathrm{sm}}$ and $\mathbf{F}_{\mathrm{mm}}$ are
the forces responsible for the secular motion and the micromotion,
respectively.  Decomposing the acceleration as $ \ddot{\mathbf{r}} = 
\ddot{\mathbf{r}}_{\mathrm{sm}} + \ddot{\mathbf{r}}_{\mathrm{mm}}$,
the micromotion being purely caused by the time-dependent part of
the potential, we have
\begin{equation}  
  m\ddot{\mathbf{r}}_{\mathrm{mm}} = -Z e \nabla_{\mathbf{r}}
  \Phi_{\mathrm{rf}}(\mathbf{r},t). 
  \label{eq:micro_mot}
\end{equation}
The equation of motion~(\ref{eq:motion}) becomes, after expansion to
first order in $\mathbf{r}_{\mathrm{mm}}$ around $\mathbf{r}_{\mathrm{sm}}$,
\begin{align} 
  m(\ddot{\mathbf{r}}_{\mathrm{sm}} + \ddot{\mathbf{r}}_{\mathrm{mm}})
  &= \mathbf{F}_{\mathrm{sm}}(\mathbf{r}_{\mathrm{sm}}) +
  \mathbf{F}_{\mathrm{mm}}(\mathbf{r}_{\mathrm{sm}},t) \nonumber \\
  &\quad + \mathbf{r}_{\mathrm{mm}} \left( \left. \nabla_{\mathbf{r}}
      \mathbf{F}_{\mathrm{sm}} 
      (\mathbf{r}) \right|_{\mathbf{r} = \mathbf{r}_{\mathrm{sm}}}
    +
    \left. \nabla_{\mathbf{r}} \mathbf{F}_{\mathrm{mm}} (\mathbf{r},t)
    \right|_{\mathbf{r} = \mathbf{r}_{\mathrm{sm}}} \right)
  \label{eq:ex_motion}
\end{align}
We now average Eq.~(\ref{eq:ex_motion}) over one period of
the micromotion, $\left(2 \pi / \Omega \right)$, and, using
Eq.~(\ref{eq:micro_mot}), we get the effective potential energy for
the time-dependent part of the potential,\cite{Major_book_2005}
\begin{align}  
  V_{\mathrm{eff},\mathrm{rf}} &= \left\langle \frac{ \left[
        \nabla_{\mathbf{r}} \Phi_{\mathrm{rf}} (\mathbf{r})
      \right]^2}{2 m} \right\rangle   \nonumber \\
  &= \frac{1}{4 m} \left( \frac{Z e V_0}{\Omega r_0^2} \right)^2
  \left( x^2 + y^2 \right)
  \label{eq:pot_eff}
\end{align}
and, consequently, the total effective potential energy is
obtained as
\begin{align} 
  V_{\mathrm{eff}} (\mathbf{r}) & = Z e \Phi_{\mathrm{s}} +
  V_{\mathrm{eff},\mathrm{rf}} \nonumber \\
  &= \left[ \frac{1}{4 m} \left( \frac{Z e V_0}{\Omega r_0^2}
    \right)^2 - \frac{Z e \kappa U_0}{2 z_0^2} \right] \left( x^2 + y^2
  \right) + \frac{Z e \kappa U_0}{z_0^2}  z^2,
  \label{eq:v_eff}
\end{align}
corresponding to a harmonic oscillator with frequencies
\begin{subequations}
  \label{eq:eff_freq}    
  \begin{align}
    \omega_x^2 = \omega_y^2 &= \left( \frac{Z e V_0}{\sqrt{2} m \Omega
        r_0^2} \right)^2 - \frac{Z e \kappa U_0}{m z_0^2}, \\
    \omega_z^2 &= \frac{2 Z e \kappa U_0}{m z_0^2}.
  \end{align}
\end{subequations}

\subsection{Floquet representation}
\label{sec:floquet}

Since the only time dependence in the Hamiltonian~(\ref{eq:hamiltoni})
is periodic with period $\Omega$, see Eq.~(\ref{eq:rf-pot}), it is
possible, using Floquet's theorem, to rewrite the solutions in terms
of an effective time-independent Hamiltonian.\cite{Shirley_PR_1965} In
the case at hand, this can lead to an analytical representation of the
wave function.  We present here a result of this Floquet analysis,
following the derivation presented in
Ref.~\refcite{Leibfried_RMP_2003}, which we will compare to our
numerical simulation with the full Hamiltonian~(\ref{eq:hamiltoni}),
and refer the reader to
Refs.~\refcite{Combescure_AIHPA_1986,Glauber_1992,Glauber_1992b,Leibfried_RMP_2003}
for more details.

Considering the problem along one dimension, one finds that the
position operator $\hat{x}$ is governed by the differential equation
\begin{equation}
  \ddot{\hat{x}} + W(t) \hat{x} = 0
\end{equation}
with
\begin{equation}
  W(t) = \frac{\Omega^2}{4} \left[ a_x + 2 q_x \cos\left( \Omega t
    \right) \right].
\end{equation}
In other words, the position operator follows the Mathieu
equation~(\ref{eq:Math.eq}), and we can write its solution as a
function $u(t)$.  For the particular initial condition
\begin{equation}
u(0) = 1; \quad \dot{u}(0) = \rmi \nu,
\end{equation}
with $\nu$ real, one finds a complete orthonormal basis set
\begin{equation}
\psi_n(t) = \rme^{-\rmi \left( n+ \frac{1}{2} \right) \nu t}
\chi_n(t),
\end{equation} 
where $\chi_n(t)$ are harmonic-oscillator-like wave functions which
ultimately depend on the solution $u(t)$ to the Mathieu equation.\cite{Leibfried_RMP_2003}
Making the approximation that $|a_x|, q_x^2 \ll 1$, one gets
\begin{equation}
u(t) \approx \rme^{\rmi \nu t } \frac{1 + \left(q_x / 2 \right) \cos
  \left( \Omega t \right)}{1+ \left(q_x / 2 \right)}
\end{equation}
with
\begin{equation}
\nu \approx \frac{\beta_x \Omega}{2}; \quad \beta_x \approx \sqrt{ a_x
  + \frac{q_x^2}{2} }.
\end{equation}
Finally, the ground state can be written as
\begin{align}
  \chi_0(t) &= \left( \frac{m \nu}{\pi \hbar} \right)^{1/4} \sqrt{
    \frac{1+ \left(q_x / 2 \right)}{1 + \left(q_x / 2 \right) \cos
      \left( \Omega t \right)}} \nonumber \\
  &\quad \times \exp\left( \left\{ \rmi \frac{m \Omega \sin \left(
          \Omega t \right)}{2 \hbar \left[ \left(2/q_x\right) + \cos
          \left( \Omega t \right) \right]} - \frac{m \nu}{2 \hbar}
    \right\} x^2 \right),
\end{align}
from which we get the time-dependent expectation value of the width as
\begin{align}
  \sigma_x &= \sqrt{ \braket{x^2} - \braket{x}^2} = \sqrt{
    \braket{x^2} } \nonumber \\
  &= \left[ \frac{\hbar}{m \Omega \left( a_x + q_x^2/2 \right)^{1/2}}
    \right]^{1/2} \left[ \frac{1+ \left(q_x / 2 \right)}{1 + \left(q_x
          / 2 \right) \cos \left( \Omega t \right)} \right]^{1/2}.
\label{eq:floquet_width}
\end{align}

One sees immediately that the norm of the approximate wave function
$\psi_0$, built from this $\chi_0$, is not conserved.  Indeed, the
pulsation of the width $\sigma_x$ is not due to the width of the
Gaussian part of $\chi_0$, but rather its amplitude.  Nevertheless, we
see that $\sigma_x$ should present an oscillation at the frequency of
the trap $\Omega$.

\section{Numerical methods}
\label{sec:num}

\subsection{Quantum mechanical approach}
\label{sec:num_quant}

In order to study the quantum dynamics of the system or, in other
words, the evolution of the wave packet of the ion trapped in a linear
Paul trap, we solve numerically\cite{Dion_CPC_2014} the
time-dependent Schr\"{o}dinger equation,
\begin{equation} 
  \rmi \hbar\frac{\partial\psi \left( t \right ) }{\partial t} =
  \hat{H}\psi (t), 
  \label{eq:shrod}
\end{equation}
with the Hamiltonian~(\ref{eq:hamiltoni}).  Starting from an initial
wave function $\psi_0 \equiv \psi(t_0)$, the solution to
Eq.~(\ref{eq:shrod}) is obtained by using time evolution
operator\cite{cohen-tannoudji92} $\hat{U}$, such that
\begin{equation} \label{eq:si}
\psi (t) = \hat{U}(t, t_{0}) \psi_0.
\end{equation}
By considering a small time increment $\Delta t$, we can use the
approximate short-time evolution operator\cite{method:pechukas66}
\begin{equation} 
  \hat{U}(t + \Delta t, t) = \exp \left\{ -\frac{\rmi}{\hbar} \int_{t}^{t
      + \Delta t} \left[ \hat{T} + \hat{V}(t') \right] \rmd t'
  \right\},
  \label{eq:unit_1}
\end{equation}
in which $\hat{T}$ and $\hat{V}(t)$ are operators corresponding to
kinetic and potential energies, respectively.

Since the time-dependent potential also has a spatial dependence,
$\hat{T}$ and $\hat{V}(t)$ do not commute and $\rme^{\hat{T} + \hat{V}}
\neq \rme^{\hat{T}} \rme^{\hat{V}}$, but a good approximation of the
evolution operator can be obtained using the split-operator
method,\cite{Dion_CPC_2014,Feit_JComputP_1982,Feit_JCP_1983}
\begin{align}  
  \hat{U}(t + \Delta t, t) &= \exp \left[ - \frac{\rmi \Delta t}{2\hbar}
    \hat{T} \right] \exp \left[ - \frac{\rmi \Delta t}{\hbar} \hat{V}(t +
    \frac{\Delta t}{2}) \right] \nonumber \\
  & \quad \times \exp \left[ - \frac{\rmi \Delta t}{2\hbar} \hat{T}
  \right] + O(\Delta t^{3}).
  \label{eq:unit_2}
\end{align}
Using a spatial grid to represent the wave function $\psi$, the above
potential energy operator can be calculated as a simple product, while
the kinetic energy operator requires the use of fast Fourier
transforms.\cite{Feit_JComputP_1982,Feit_JCP_1983} More details on the
numerical algorithm used here can be found in
Ref.~\refcite{Dion_CPC_2014}.  This approach to the time evolution is
more demanding numerically than, for example, the method used in
Ref.~\refcite{Li_PRA_1993}, where only three coupled equations were 
solved by a Runge-Kutta method to get the time-dependent evolution
operator.  The latter requires nevertheless the use of a basis set of
functions to recover the wave function whenever observables need to be
calculated.  The main advantage of our approach is that it is not
constrained to a particular form of the Hamiltonian, and can be easily
extended in the future to allow the simulation of molecular ions or
interactions with laser pulses.

In order to compare the quantum and classical trajectories, we need an
initial wave function that will not spread out during the time
evolution.  This is possible by using a coherent
state,\cite{Leibfried_RMP_2003,Glauber_1992b} which corresponds to a
displaced ground-state wave function.  We thus start from the
stationary solutions to the Schr\"{o}dinger equation with the effective
potential $V_{\mathrm{eff}}$ [Eq.~(\ref{eq:v_eff})],
\begin{equation} 
  \psi_{n_x n_y n_z}(x,y,z) = \phi_{n_x}(x) \phi_{n_y}(y)
  \phi_{n_z}(z),
  \label{eq:oscillator3D}
\end{equation}
where the $\phi_n$ are solutions to the one-dimensional harmonic
oscillator,\cite{Bransden_Joachain_2003}
\begin{equation}
  \phi_{n}(\xi) = \left(\frac{1}{\sigma_{0\xi} \sqrt{\pi} 2^{n} n!}
  \right)^{\frac{1}{2}} \exp \left( -\frac{\xi^{2}}{2 \sigma_{0\xi}^2}
  \right) H_{n}\left(\frac{\xi}{\sigma_{0\xi}}\right), 
  \label{eq:oscillator1D}
\end{equation}
where 
\begin{equation}
  \sigma_{0\xi} \equiv \left( \frac{\hbar^{2}}{m^2 \omega_{\xi}}
  \right)^{\frac{1}{4}},
  \label{eq:sigma0}
\end{equation}
$H_{n}$ is the Hermite polynomial, and $\xi \in \{ x, y, z\}$.
Instead of displacing the ground state, we give the ion an initial
momentum $\mathbf{p}_0 = \hbar \mathbf{k}_0$, we add a complex phase
factor to the ground state wave function, i.e., we use as the initial
wave function
\begin{equation}
  \psi(\mathbf{r}, t=0) = \phi_0(x) \phi_0(y) \phi_0(z) \rme^{\rmi
    \mathbf{k}_0 \cdot \mathbf{r}}.
  \label{eq:initial state}
\end{equation}
(We have also performed simulations, not presented in this paper,
where the initial momentum was set to zero and the wave packet was
instead displaced initially from the center of the trap.  The results
obtained were qualitatively the same as those reported here.)

Finally, it should be noted that with this choice for the initial wave
function, together with the (time-dependent) potential and time
evolution operator, the problem is separable in the different spatial
coordinates, and the solution of the 3D Schr\"{o}dinger equation can be
reduced to a superposition of 1D problems, \ie,
\begin{equation}
\psi(\mathbf{r}, t) = \psi_x(x,t) \psi_y(y,t) \psi_z(z,t).
\end{equation}

\subsection{Wave function in the effective potential approximation}
\label{sec:effective}

Considering the Schr\"{o}dinger equation in Eq.~(\ref{eq:shrod}),
following Ref.~\refcite{Major_book_2005} we can write a general solution
to this equation as
\begin{equation} 
  \psi (t) = \exp \left[ - \frac{\rmi}{\hbar} W(t) \right] \varphi(t),
  \label{eq:actual_si}
\end{equation}
where $W$ is a function of space coordinates and time such that
\begin{equation} \label{eq:w_1}
  \frac{\partial W(t)}{\partial t} = V(t),
\end{equation} 
in which $V(t)$ is the time-dependent part of the potential energy
corresponding to the potential energy in Eq.~(\ref{eq:rf-pot}). In our
case, $W$ has the form
\begin{equation} \label{eq:w_2} W(t) = \frac{1}{2}\frac{QV_{0}}{\Omega
    r_{0}^{2}}(x^{2} - y^{2})\sin(\Omega t),
\end{equation}
and the time average of $W$ is zero.
In these conditions, a good approximation to the actual wave function
can be obtained using 
\begin{equation} \label{eq:si_eff}
  \psi_{\mathrm{eff}} (t) = \exp \left[ - \frac{\rmi}{\hbar} W(t) \right]
  \phi_{\mathrm{eff}}(t),
\end{equation}
where $\phi_{\mathrm{eff}}(t)$ is the wave function obtained from the
time evolution of the initial state using the effective potential
Eq.~(\ref{eq:pot_eff}) instead of the actual time-dependent potential,
Eq.~(\ref{eq:pot}).

There is a limit for the validity of the effective potential
approximation which requires $|a|, |q| \ll 1$.\cite{Major_book_2005} In
this limit the effective wave function in Eq.~(\ref{eq:si_eff}) can be
a good approximation of the real wave function. We will compare here
the actual wave function with the effective one for different sets of
the values $(a, q)$.

\subsection{Classical approach}
In order to compare the quantum motion of the ion with its classical
approximation, we need to employ a numerical method for the
integration of the classical equations of
motion~(\ref{eq:eq_motion_3D}).  In particular, we wish to conserve
the symplectic flow of the Hamiltonian system and therefore take the
St\"{o}rmer-Verlet scheme as a symplectic
integrator.\cite{Hairer_book_2006}

To second order in time, the evolution of the dynamical system is
given by 
\begin{align*} 
  \mathbf{p}_{n + \frac{1}{2}} &= \mathbf{p}_{n} - \frac{\Delta t}{2}
  \mathbf{H}_{\mathbf{r}} \left( \mathbf{p}_{n + \frac{1}{2}},
    \mathbf{r}_{n} \right), \nonumber \\ 
  \mathbf{r}_{n + 1} &= \mathbf{r}_{n} + \frac{\Delta t}{2} \left[
    \mathbf{H}_{\mathbf{p}} \left( \mathbf{p}_{n +
        \frac{1}{2}}, \mathbf{r}_{n} \right ) + \mathbf{H}_{\mathbf{p}}
    \left( \mathbf{p}_{n + \frac{1}{2}}, 
      \mathbf{r}_{n + 1} \right) \right], \nonumber \\
  \mathbf{p}_{n + 1} &= \mathbf{p}_{n + \frac{1}{2}} - \frac{\Delta
    t}{2} \mathbf{H}_{\mathbf{r}} \left( \mathbf{p}_{n + \frac{1}{2}},
    \mathbf{r}_{n + 1} \right), 
\end{align*}
where $\mathbf{H}_{\mathbf{p}}$ and $\mathbf{H}_{\mathbf{r}}$ are
column vectors of partial derivatives of the Hamiltonian
$H(\mathbf{p},\mathbf{q})$ with respect to the components of the
momentum $\mathbf{p}$ and position $\mathbf{r}$, and $\Delta t$ stands
for the step size of the time steps indexed by $n$.
For our Hamiltonian, this system reduces to
\begin{subequations}
  \label{eq:StormerVerlet}
  \begin{align} 
    \mathbf{p}_{n + \frac{1}{2}} &= \mathbf{p}_{n} - \frac{\Delta t}{2}
    \mathbf{H}_{\mathbf{r}_n}, \\ 
    \mathbf{r}_{n + 1} &= \mathbf{r}_{n} + \Delta t
    \frac{\mathbf{p}_{n + \frac{1}{2}}}{m}, \\ 
    \mathbf{p}_{n + 1} &= \mathbf{p}_{n + \frac{1}{2}} - \frac{\Delta
      t}{2} \mathbf{H}_{\mathbf{r}_{n+1}}, 
  \end{align}
\end{subequations}
where the components of $\mathbf{H}_{\mathbf{r}_{n}}$ are given by
\begin{subequations}
  \begin{align}
    H_{x_n} &= Ze \left[ -\frac{\kappa U_0}{z_0^2} + \frac{V_0}{r_0^2} \cos (
      \Omega t_n) \right] x_n, \\
    H_{y_n} &= Ze \left[ -\frac{\kappa U_0}{z_0^2} - \frac{V_0}{r_0^2} \cos (
      \Omega t_n) \right] y_n, \\
    H_{z_n} &= \frac{2 Z e \kappa U_0}{z_0^2}  z_n.
  \end{align}
\end{subequations}
The system of equations~(\ref{eq:StormerVerlet}) is iterated starting
from an initial condition equivalent to Eq.~(\ref{eq:initial state}),
namely $\mathbf{r_0} = \mathbf{0}$ and $\mathbf{p}_0 = \hbar \mathbf{k}_0$.

\section{Results}

For our simulations, we use the atomic ion Ca$^+$ as an example. All
the trapping configuration parameters are kept constant for all
simulations, with the exception of the amplitudes of static and
radio-frequency electric potentials $U_{0}$ and $V_{0}$. The value of
the fixed parameters are given in Tab.~\ref{tab:param}, with the trap
parameters based on typical experimental
realizations.\cite{Raizen_PRA_1992}
\begin{table}
  \tbl{Trap configuration parameters for a Ca$^+$ ion.}
  {
    \begin{tabular}{cc}
      \hline \hline
      Parameter & Value \\ \hline
      $m$ & \SI{6.6529e-26}{kg} \\
      $r_{0}$ &  \SI{0.769e-3}{m} \\
      $z_{0}$ &  \SI{1.25e-3}{m} \\
      $\kappa$ &  $0.31$ \\
      $\Omega$ &   $2\pi \times \SI{8e6}{\per\second}$ \\
      \hline \hline
    \end{tabular}
  \label{tab:param}}
\end{table}

Unless noted otherwise, the numerical simulations were run using the
following parameters.  We have taken time steps $\Delta t =
\SI{1e-10}{s}$ for both quantum and classical simulations, for a total
simulation time of \SI{5}{\micro\second}.  For the voltages used, the
trapping potential is more elongated in the $z$ direction, such that the
spatial grid for the quantum simulations spans $[
\SI{-1}{\micro\meter}, \SI{1}{\micro\meter}]$ in $x$ and $y$ and $[
\SI{-3}{\micro\meter}, \SI{3}{\micro\meter}]$ along $z$, with 2000 grid
points along each direction.  The initial momentum of the ion is set
to $p_0 = \SI{6.777e-26}{\kilo\gram\meter\per\second}$, which
corresponds approximately to the Doppler temperature achieved by laser
cooling of Ca$^+$, namely $\SI{4.4}{\milli\kelvin}$.\cite{Urabe_APB_1998}

\subsection{Quantum vs Classical Trajectories}
\label{sec:QvsC}

We ran simulations, classically and quantum mechanically, for three
different pairs of $\left( U_{0}, V_{0} \right)$ equal to $(\SI{2}{V},
\SI{50}{V})$, $(\SI{8}{V}, \SI{90}{V})$ and $(\SI{10}{V},
\SI{140}{V})$. All these values result in pairs of $(a,q)$ which are
located inside the stability region, see Fig.~\ref{fig:stability}, and
therefore these values are expected to form bounded trajectories for
the motion of the ion inside the trap.

\begin{figure}
\includegraphics[width=0.49\textwidth]{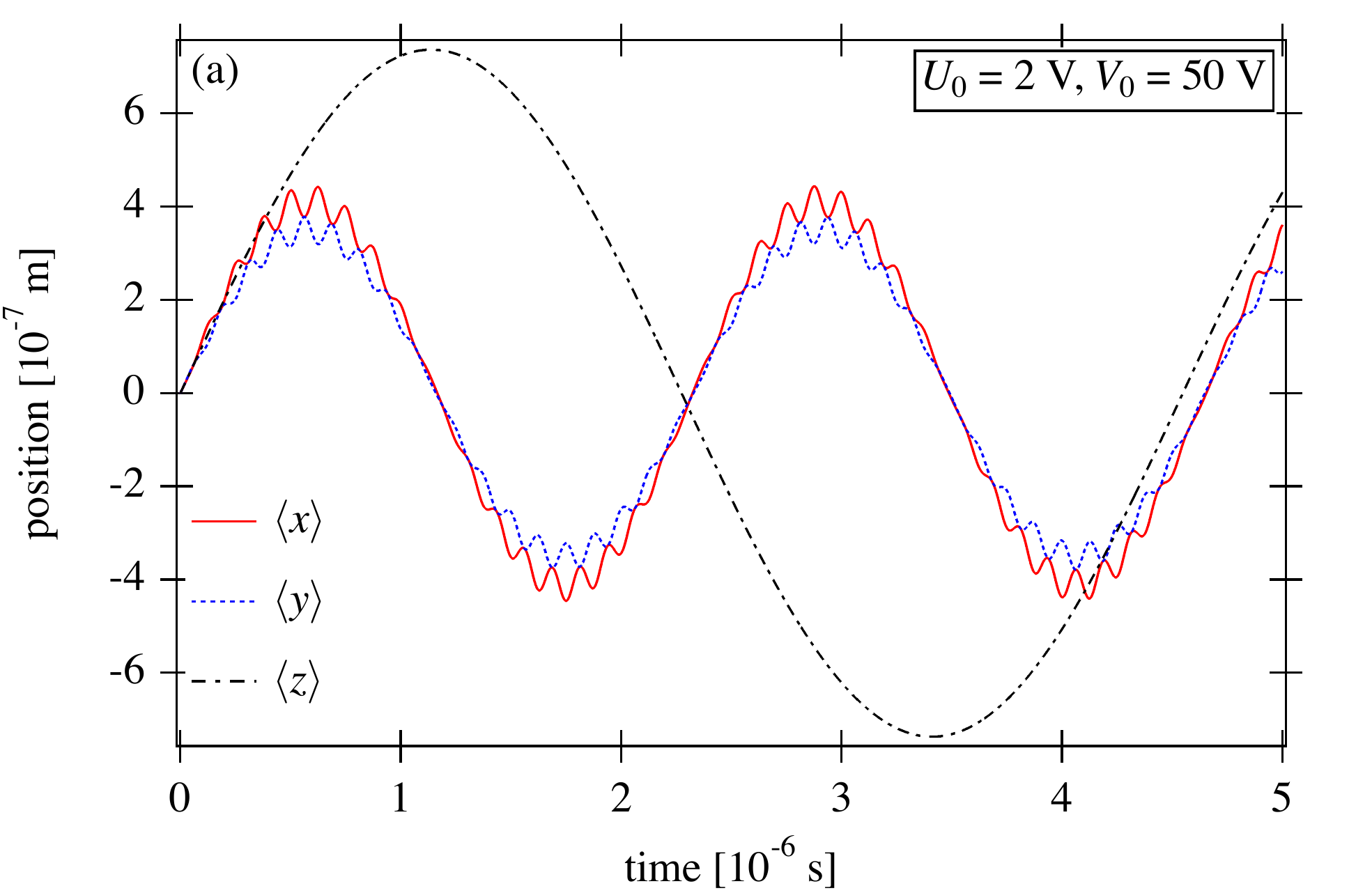}   
\includegraphics[width=0.49\textwidth]{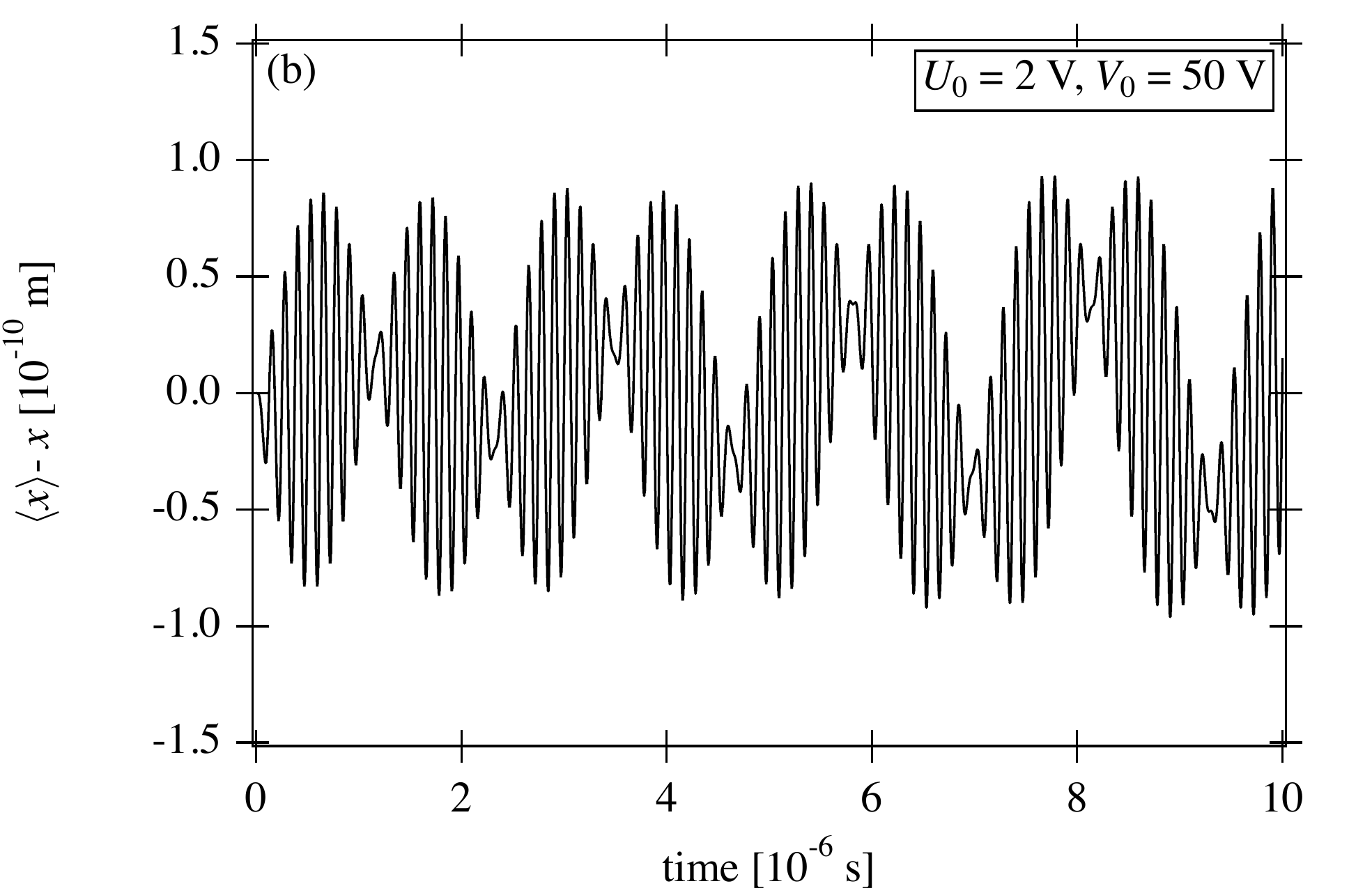} \\
\includegraphics[width=0.49\textwidth]{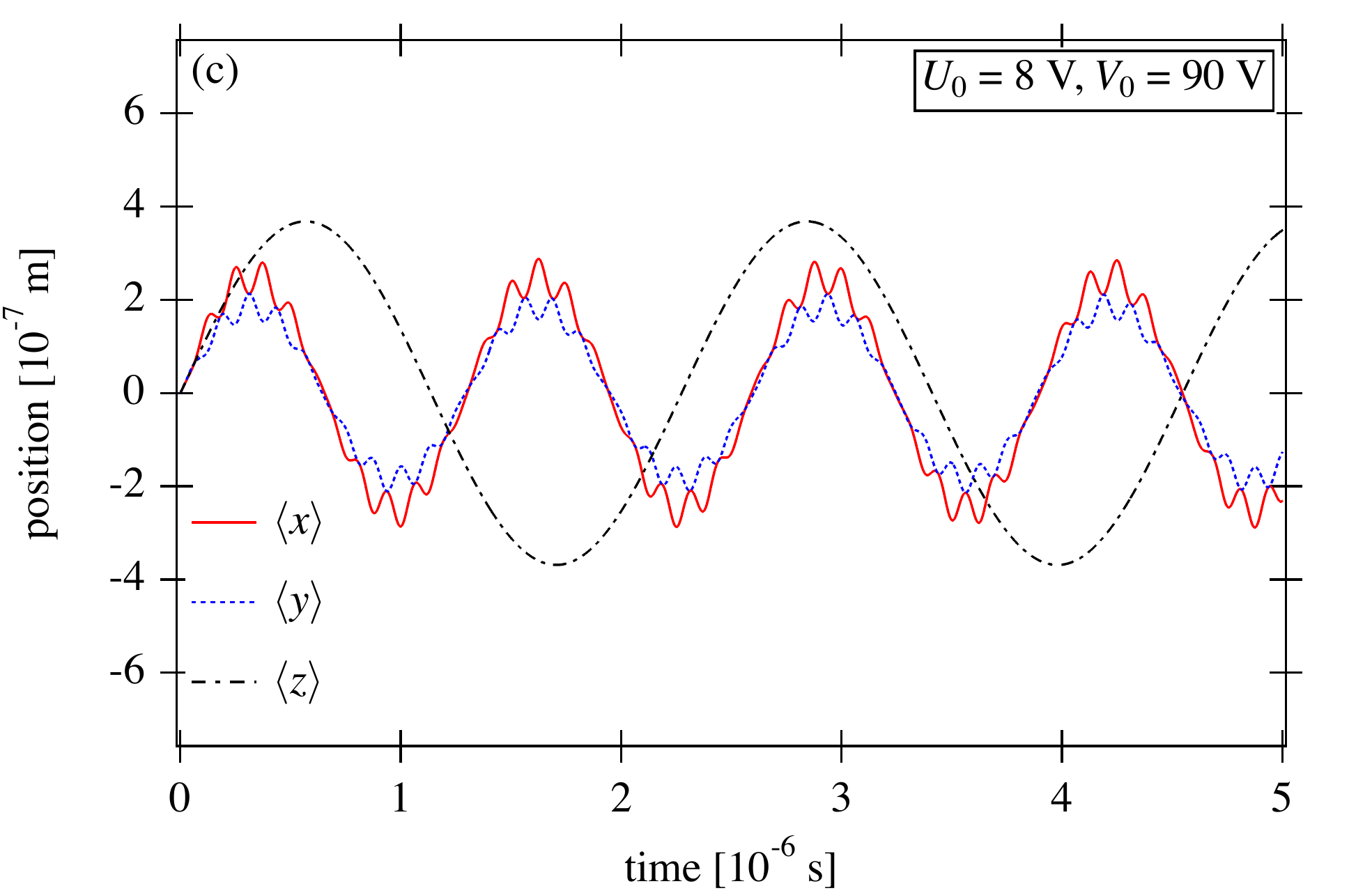}    
\includegraphics[width=0.49\textwidth]{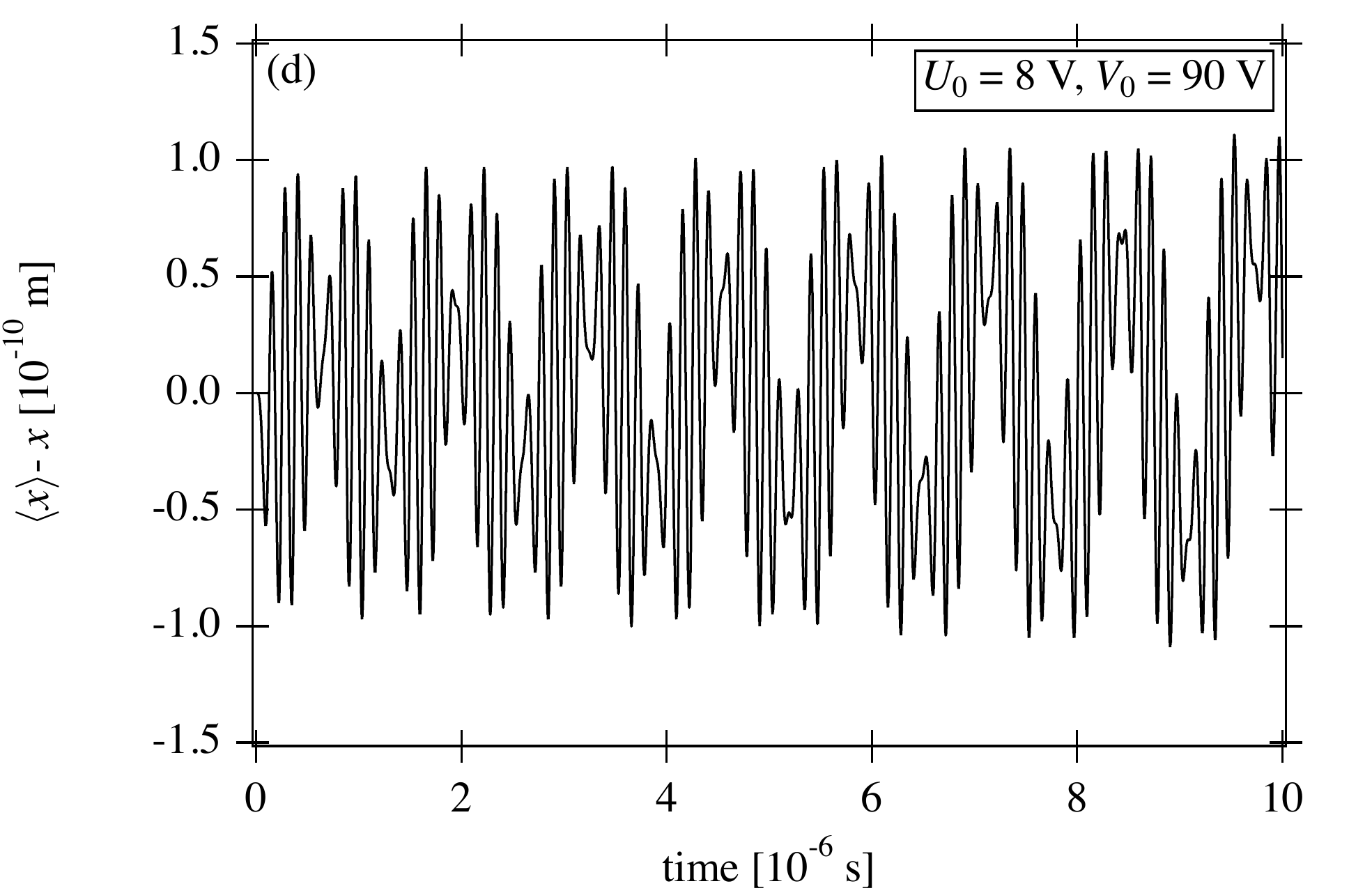} \\
\includegraphics[width=0.49\textwidth]{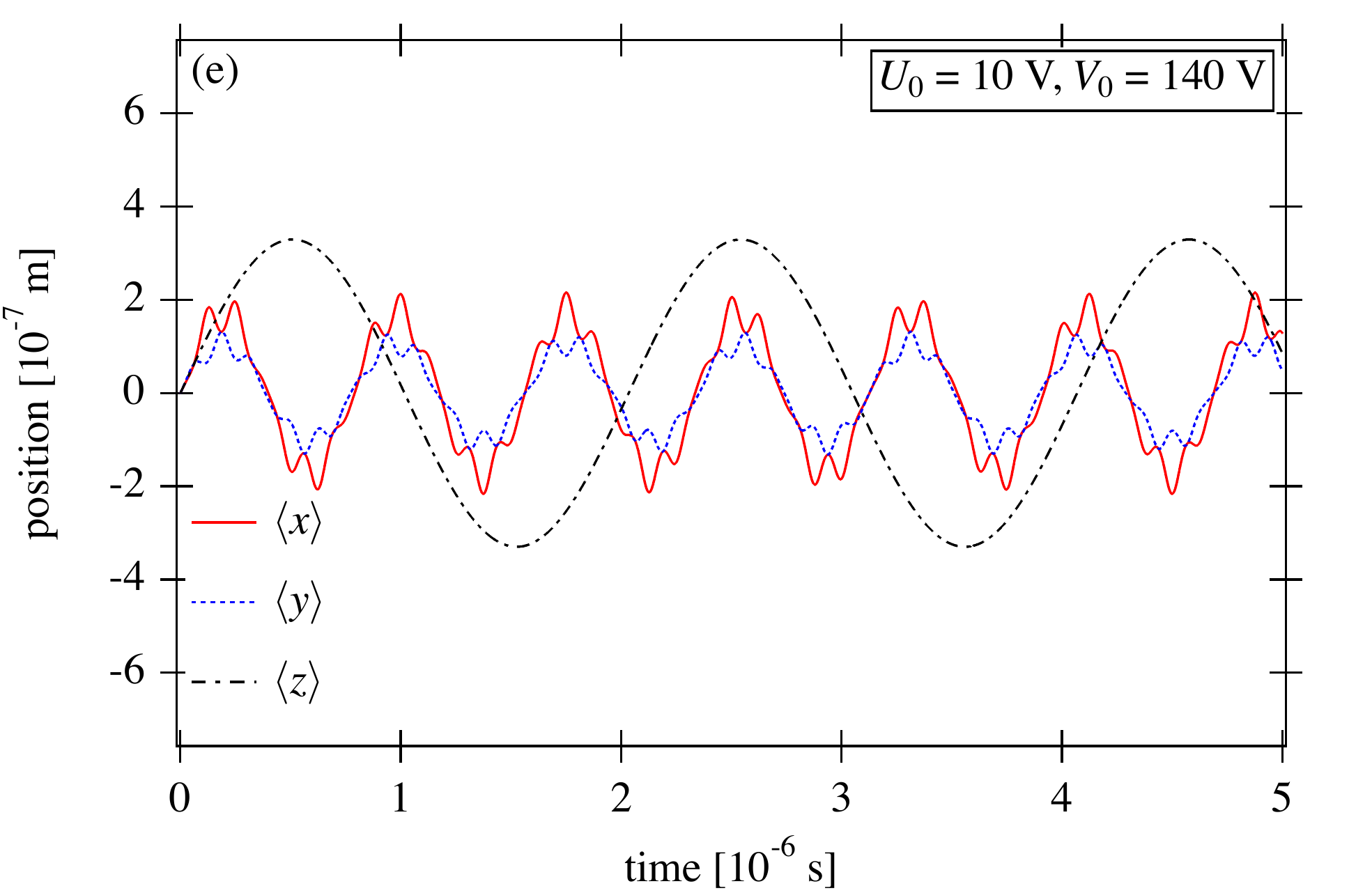}    
\includegraphics[width=0.49\textwidth]{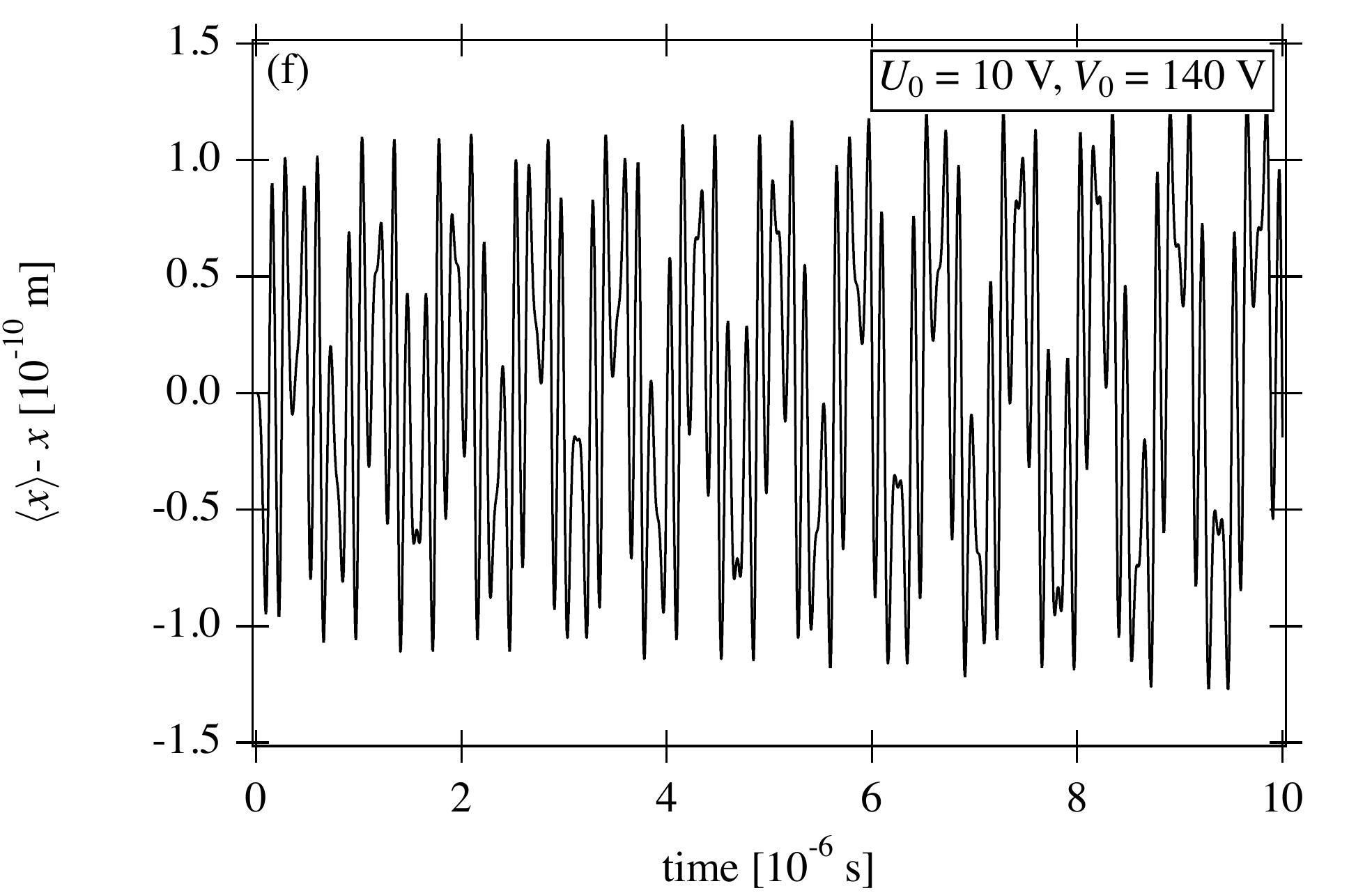}
\caption{\label{fig:evolution}(Color online) Time evolution of the
  center-of-mass motion for different trap potentials: (a)--(b) $U_0 =
  \SI{2}{V}$, $V_0 = \SI{50}{V}$; (c)--(d) $U_0 = \SI{8}{V}$, $V_0 =
  \SI{90}{V}$; (e)--(f) $U_0 = \SI{10}{V}$, $V_0 = \SI{140}{V}$.
  Panels (a), (c), and (e) present the components of the expectation
  value of the center of mass, $\braket{\mathbf{r}}$ for the quantum
  trajectory.  Panels (b), (d), and (f) present the difference between
  the quantum-mechanical expectation value $\braket{x}$ and the
  classical trajectory $x$.}
\end{figure}

The results obtained are shown in Fig.~\ref{fig:evolution}, where the
$x$, $y$, and $z$ components of the center-of-mass motion are shown in
panels (a), (c), and (e).  
For all three sets of parameters, we clearly see that $\braket{x}$ and
$\braket{y}$ present an overall harmonic motion, on which micromotion
is superposed, as expected (see Sec.~\ref{sec:effpot}).  By contrast,
the motion is harmonic in $z$, which is governed only by the ring-shaped
electrodes, on which a static potential is applied.  The frequency of
the secular, harmonic motion also follows the dependence on $U_0$ and
$V_0$ obtained form the classical model, Eqs.~(\ref{eq:eff_freq}).  A
direct comparison between the quantum and classical trajectories is
presented in Fig.~\ref{fig:evolution}(b), (d), and (f), for the $x$
component of the center-of-mass motion.  We note that the absolute
error is smaller than the grid spacing of the quantum simulation,
$\Delta x = \SI{e-9}{m}$, but most importantly, it does not
significantly grow with time, indicating that the periodicity of the
motion is indeed the same.

\subsection{Width of the wave packet}
\label{sec:width}

\subsubsection{Spreading of the wave packet}
\label{sec:spreading}

\begin{figure}
  \begin{center}
    \includegraphics[width=0.65\textwidth]{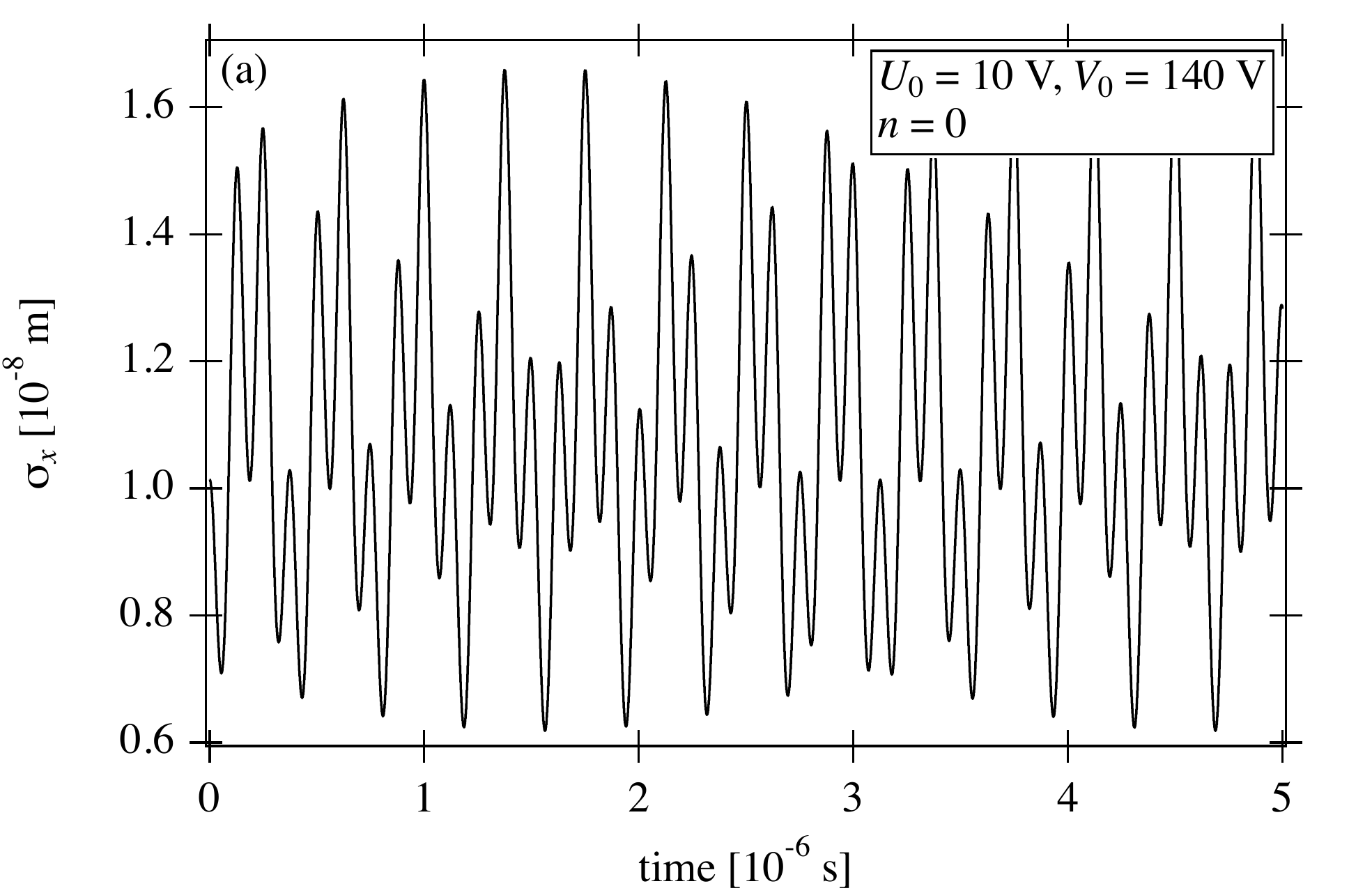}
    \includegraphics[width=0.65\textwidth]{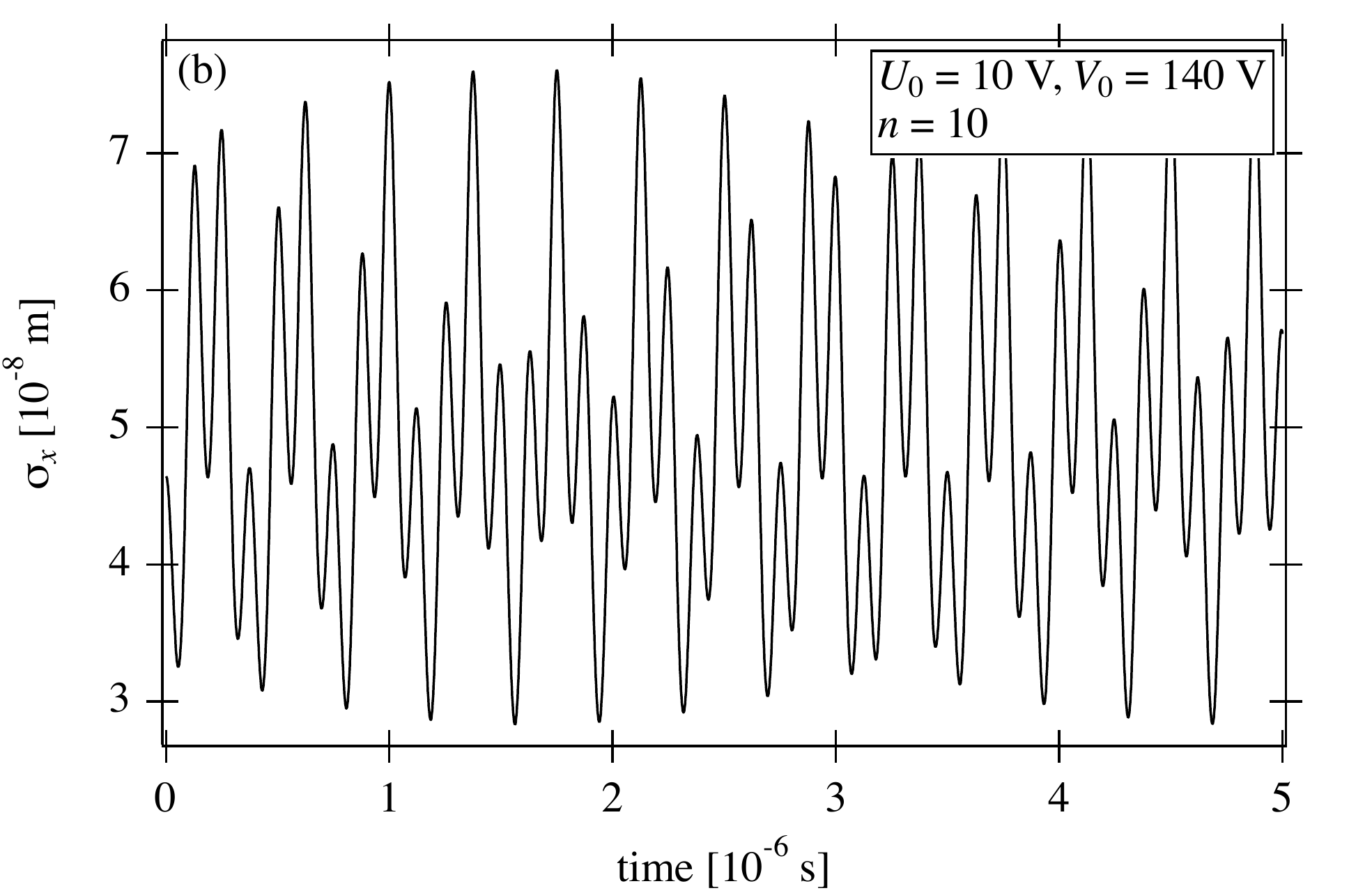}
  \end{center}
  \caption{\label{fig:width}Width $\sigma_x$ of the
    wave packet, Eq.~(\ref{eq:width}), for $U_{0} = \SI{10}{V}$, $V_{0}
    = \SI{140}{V}$, starting from the initial state: (a) Gaussian
    ground state of the effective potential [$n=0$ in
    Eq.~(\ref{eq:oscillator1D})]; (b) excited state of the effective
    potential [$n=10$ in Eq.~(\ref{eq:oscillator1D})].}
\end{figure}

One information that is not available from the classical simulation is
the width of the wave packet and its possible growth. We have thus
calculated, from the quantum simulations, the width of the wave
packet, according to
\begin{equation}
\sigma_{\xi} \equiv \sqrt{\braket{\xi^2} - \braket{\xi}^2},
\label{eq:width}
\end{equation}
with $\xi \in \{x,y,z\}$. A sample result, for $\left( U_{0}, V_{0}
\right) = (\SI{10}{V}, \SI{140}{V})$, is shown in
Fig.~\ref{fig:width}.
We see that the width of the wave packet oscillates with time, but
with that oscillation bounded, such that the ion remains completely
inside the trap.  This behavior is expected as the initial state
chosen, see Sec.~\ref{sec:num_quant}, corresponds to a Gaussian
coherent state, which were previously shown to result in periodic
oscillations of the width of the Gaussian.\cite{Glauber_1992b}

\begin{figure}
  \begin{center}
    \includegraphics[width=0.65\textwidth]{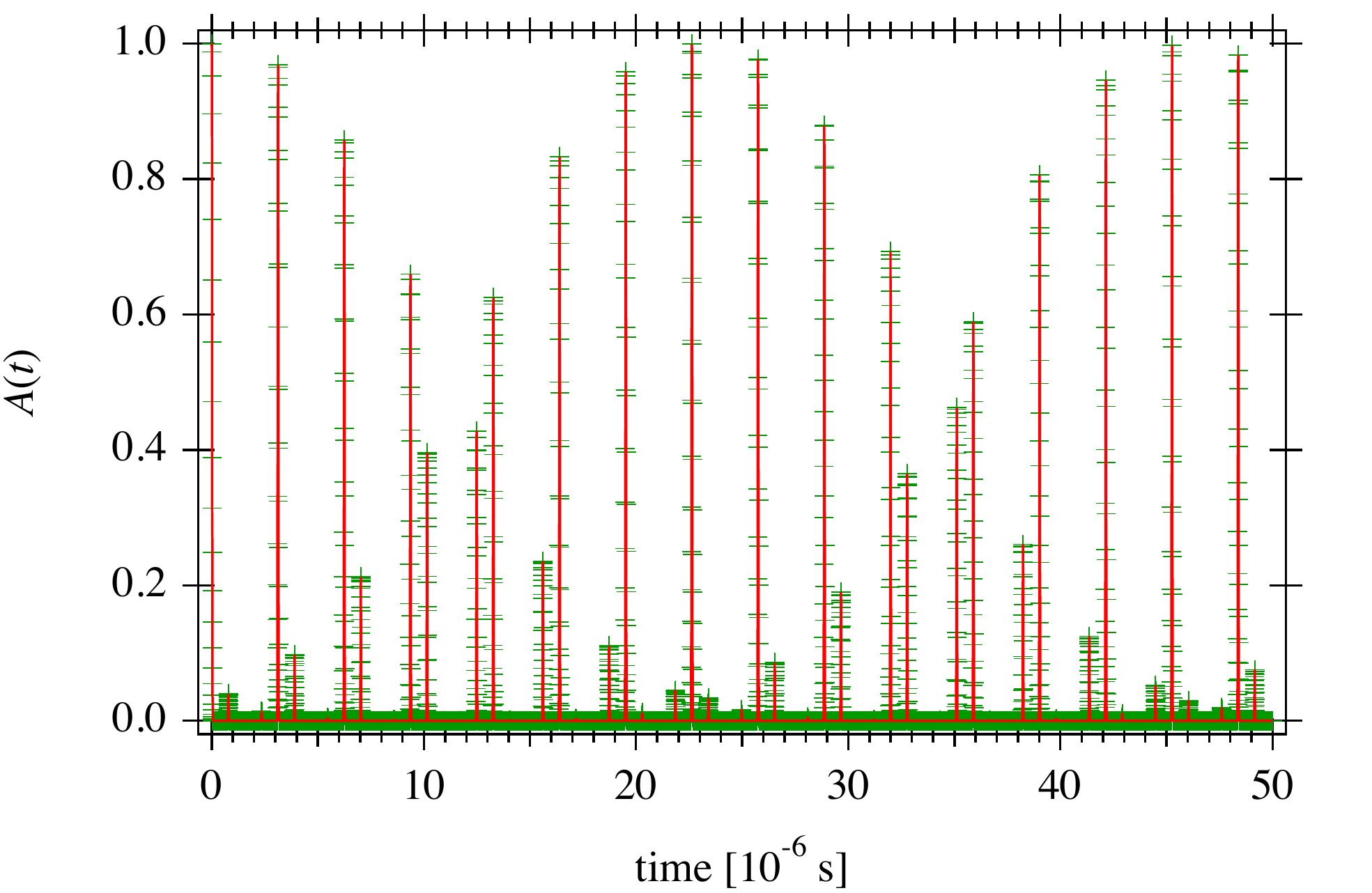}
  \end{center}
  \caption{\label{fig:autoc}(Color online) Autocorrelation function
    for $U_{0} = \SI{10}{V}$, $V_{0} = \SI{140}{V}$, starting from:
    the initial state Gaussian ground state of the effective potential
    [$n=0$ in Eq.~(\ref{eq:oscillator1D})] (crosses); the excited state
    of the effective potential [$n=10$ in Eq.~(\ref{eq:oscillator1D})]
    (full line).}
\end{figure}

We have also done simulations for cases where the ion is not initially
in the ground state of the effective potential [$n \neq 0$ in
Eq.~(\ref{eq:oscillator3D})], and thus obviously not a coherent
state.\cite{Klauder_book_1985}  We find that such a situation leads
to exactly the same motion of the center of mass.  What is more
striking is that the oscillation of the wave packet also presents the
same behavior in both cases, see Fig.~\ref{fig:width}.  Except for the
amplitude of the oscillation, we see that the two cases follow the
same temporal variation.  This is evidenced also by the
autocorrelation function
\begin{equation}
  A_x(t) = \left| \Braket{\psi_x(x,0) | \psi_x(x,t)} \right|^2,
  \label{eq:autocorrelation}
\end{equation}
which is plotted in Fig.~\ref{fig:autoc} for both $n=0$ and $n=10$,
which show a return to the initial state after \SI{2.26e-5}{s} (we
have checked that this result is independent of the choice of $n$).
This recurrence, which depends on the parameters $U_0$ and $V_0$,
corresponds to the time it takes for the classical trajectory to
return to the same point in phase space, reinforcing the link between
the classical and quantum trajectories, see
Sec.~\ref{sec:stability_width}.  Due to the symmetry in the trapping
field, see Eqs.~(\ref{eq:pot})--(\ref{eq:st-pot}), this recurrence
time is also observed for $\psi_y(y,t)$.  However, along the $z$ axis
the ion has a simple harmonic motion, with a period of
\SI{2.03e-6}{s}, so the full autocorrelation function
\begin{equation}
  A(t) = \left| \Braket{\psi(0) | \psi(t)} \right|^2,
\end{equation}
doesn't show any return to $A=1$.  

\subsubsection{Stability and spreading of the wave packet}
\label{sec:stability_width}

Previous work has established that the motion of the center of mass of
the wave packet should follow the same trajectory as that for a
classical particle.\cite{Combescure_AIHPA_1986,Brown_PRL_1991} As
such, the stability criterion derived from the Mathieu equation, see
Sec.~\ref{sec:stability}, should apply also to the quantum
dynamics,\cite{Li_PRA_1993,Combescure_AIHPA_1986} provided also that
the oscillation of the width of the wave packet stays bounded, see
Sec.~\ref{sec:spreading}.

\begin{figure}
\includegraphics[width=0.49\textwidth]{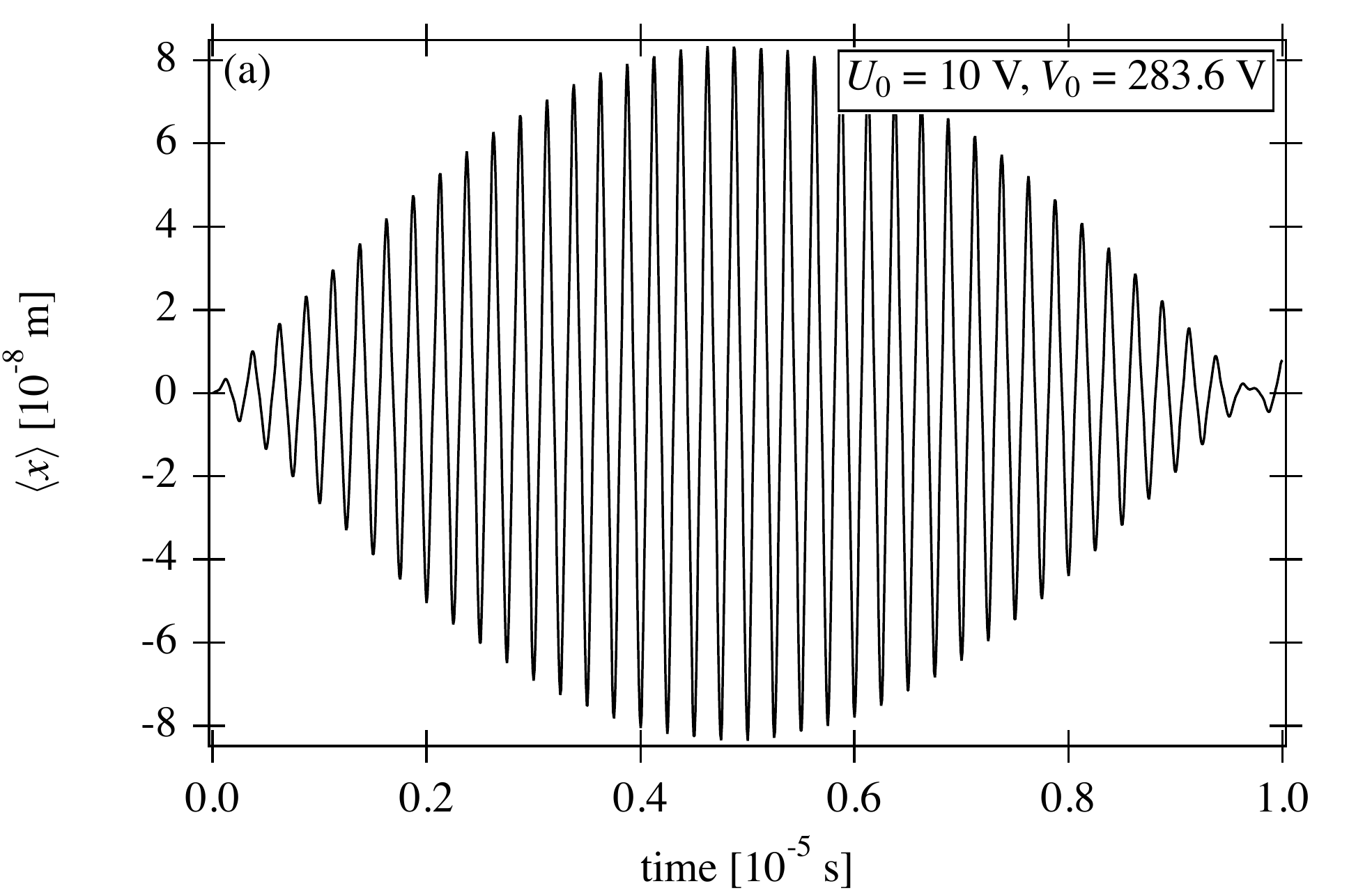}
\includegraphics[width=0.49\textwidth]{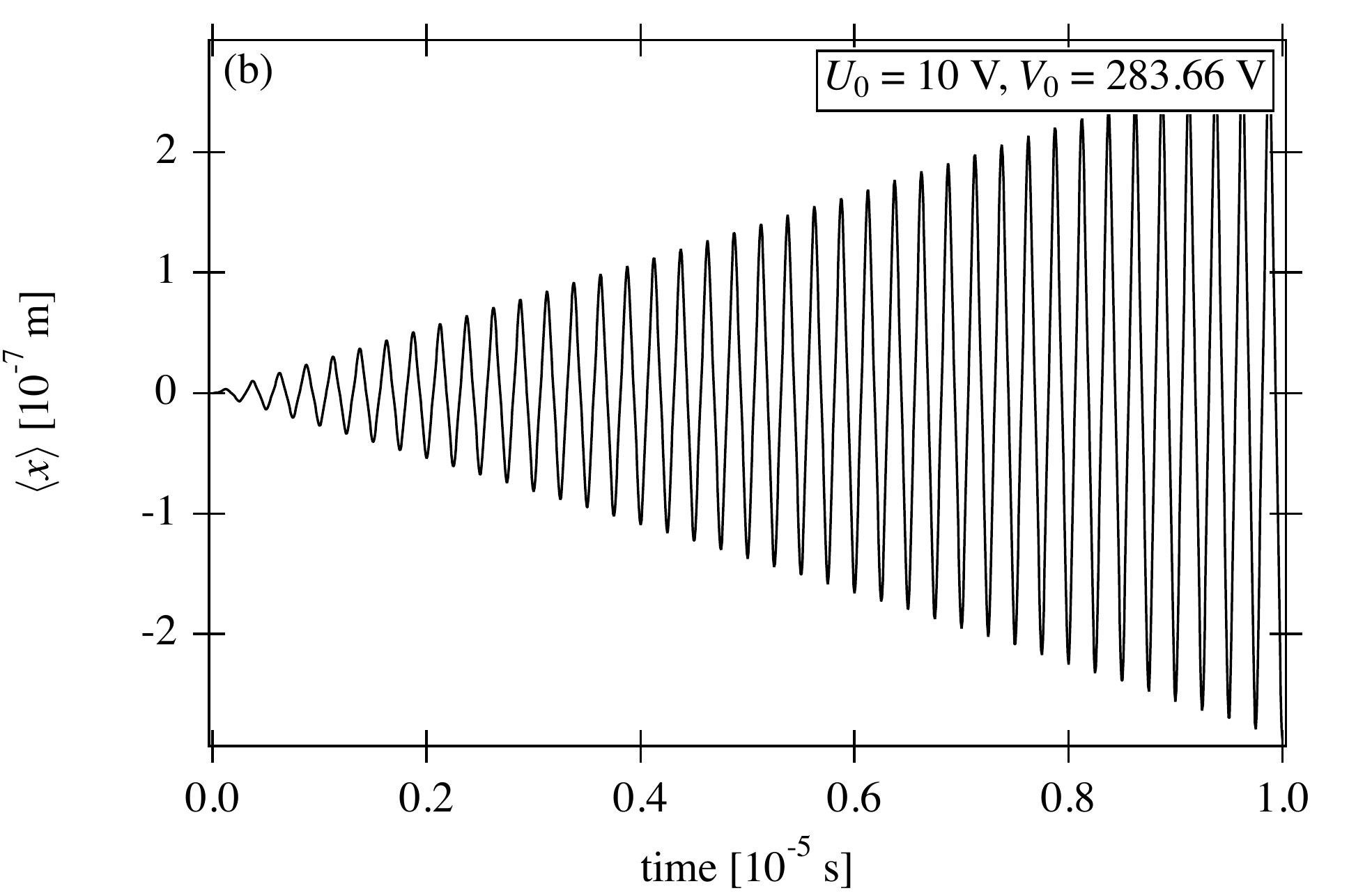} \\
\includegraphics[width=0.49\textwidth]{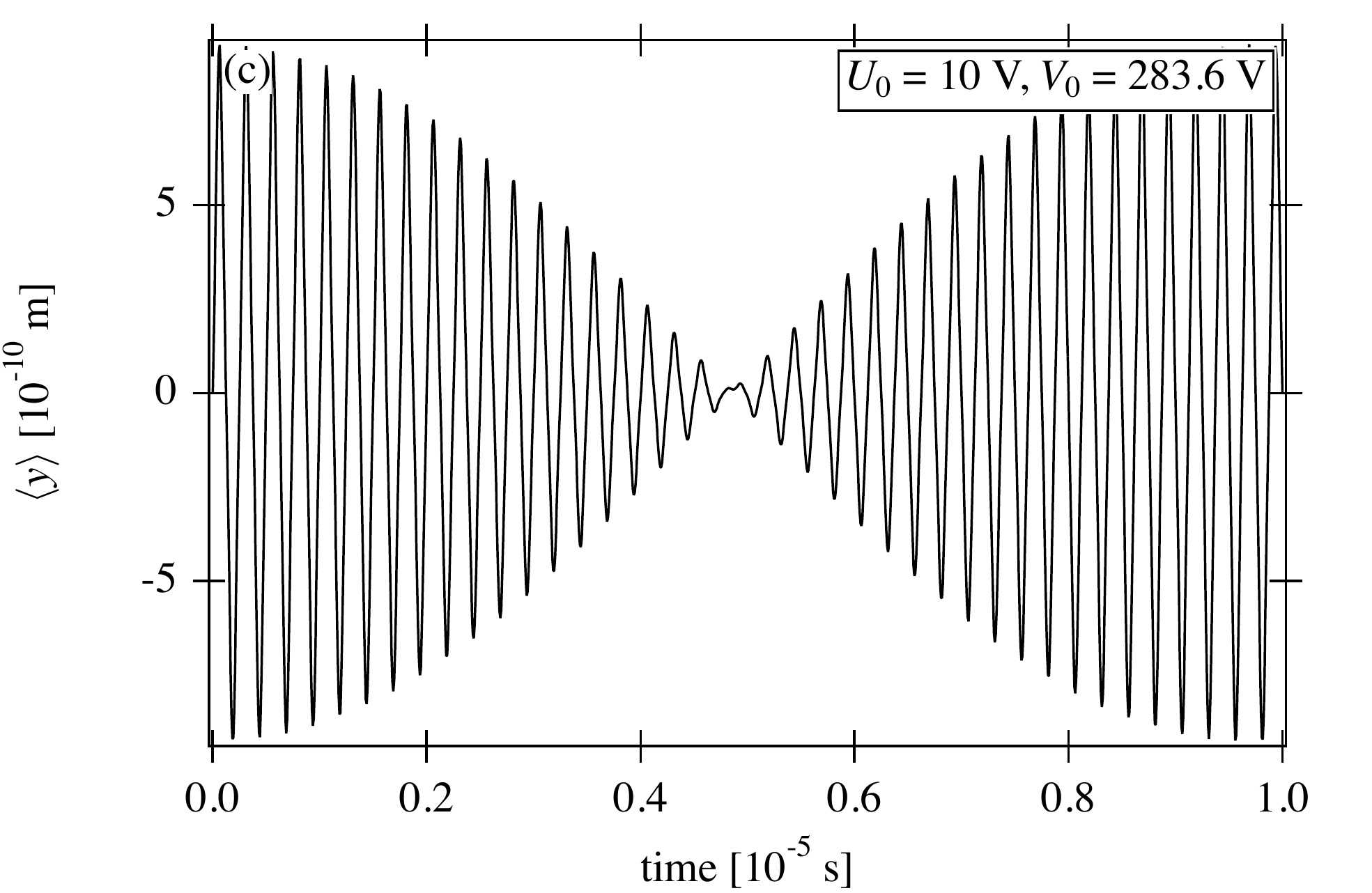}
\includegraphics[width=0.49\textwidth]{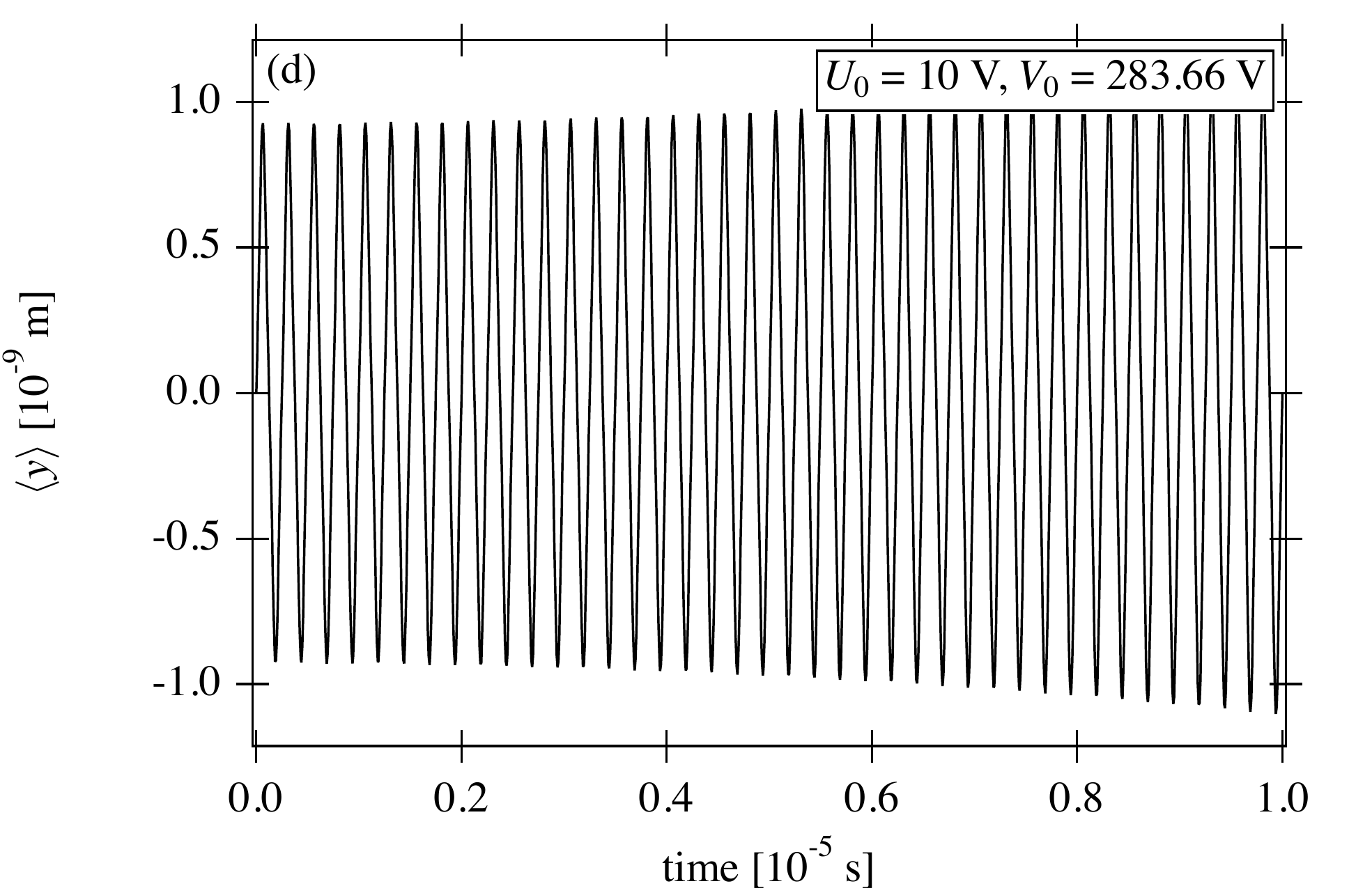} \\
\includegraphics[width=0.49\textwidth]{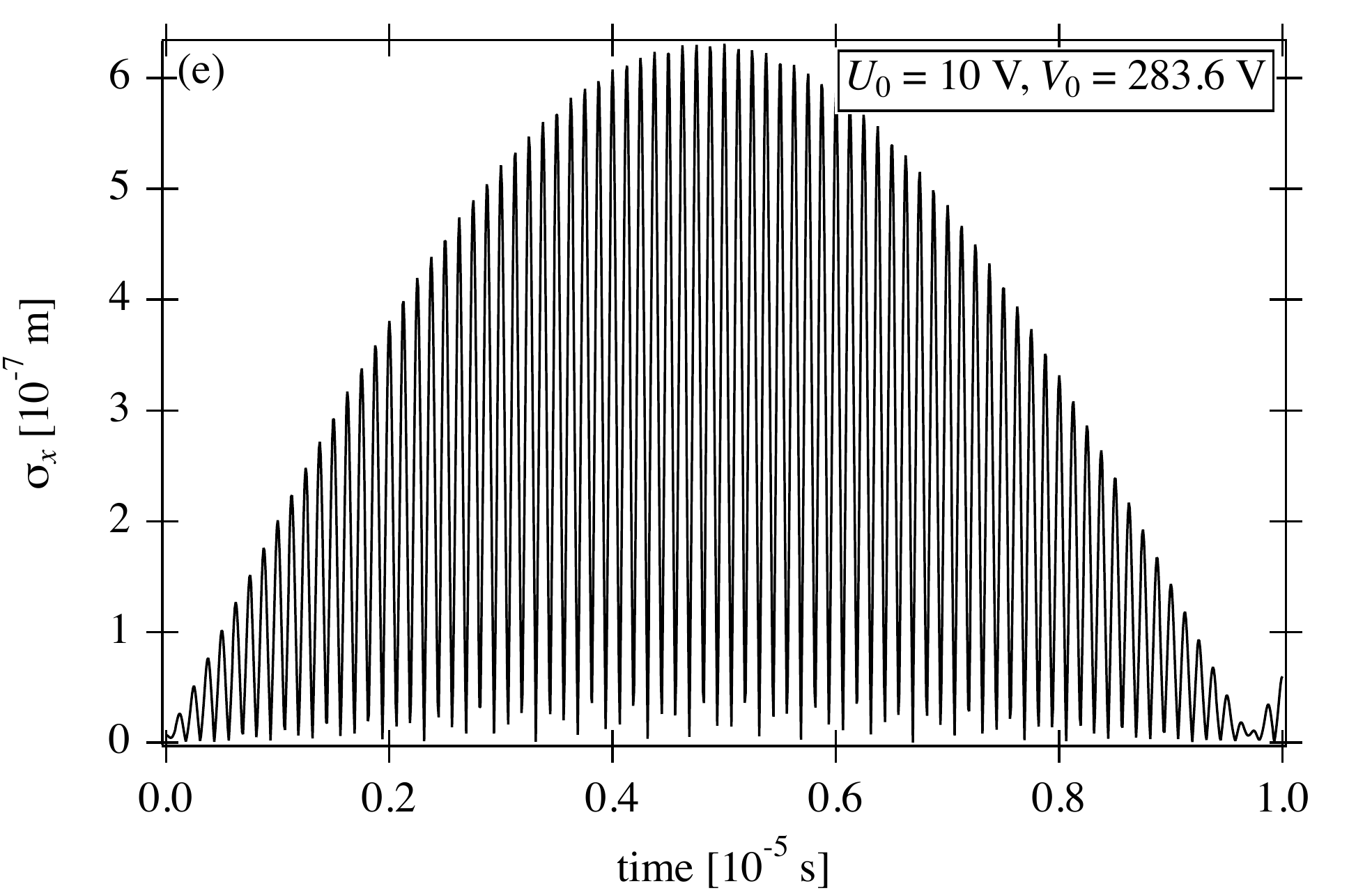}
\includegraphics[width=0.49\textwidth]{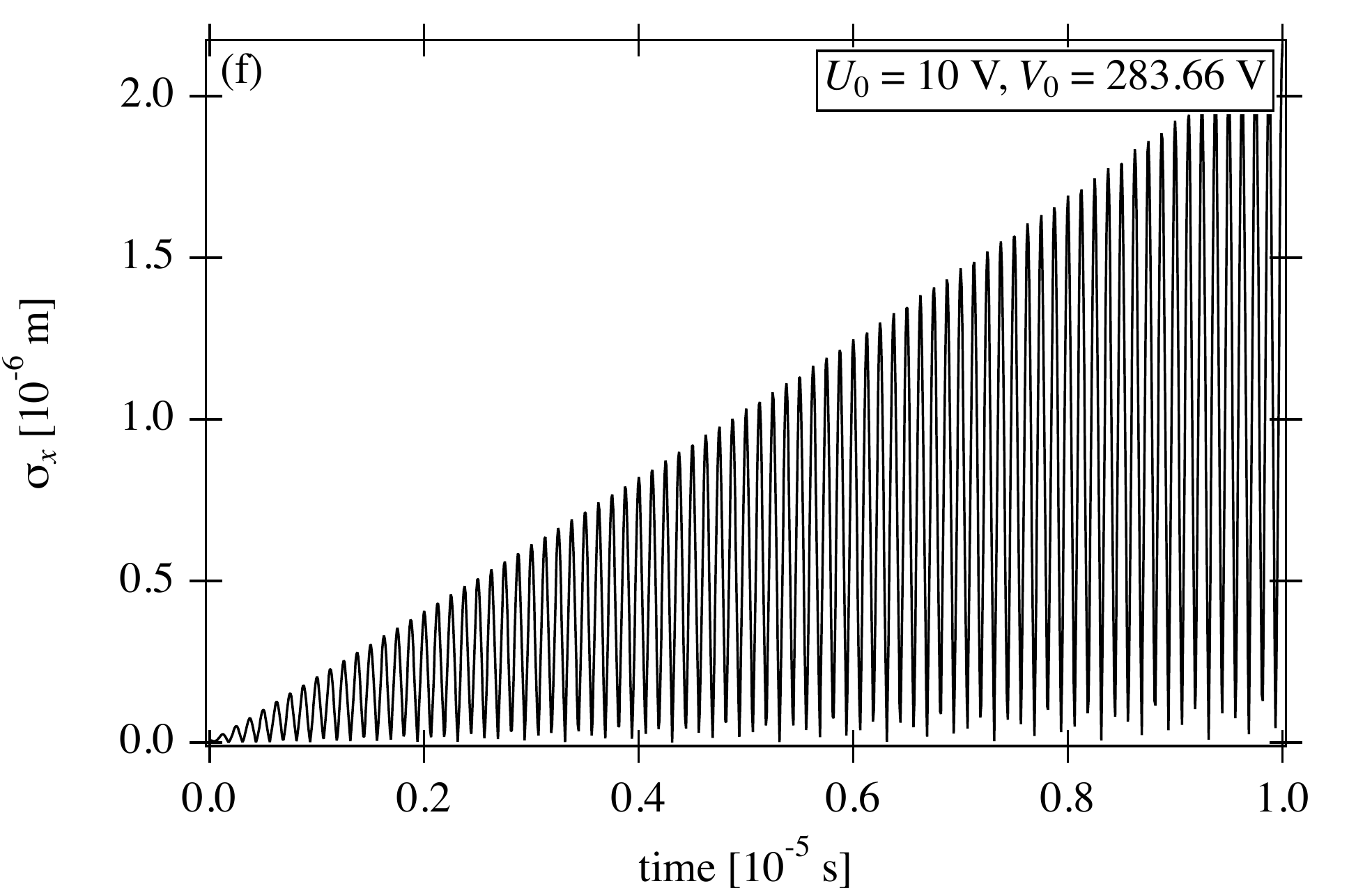} \\
\includegraphics[width=0.49\textwidth]{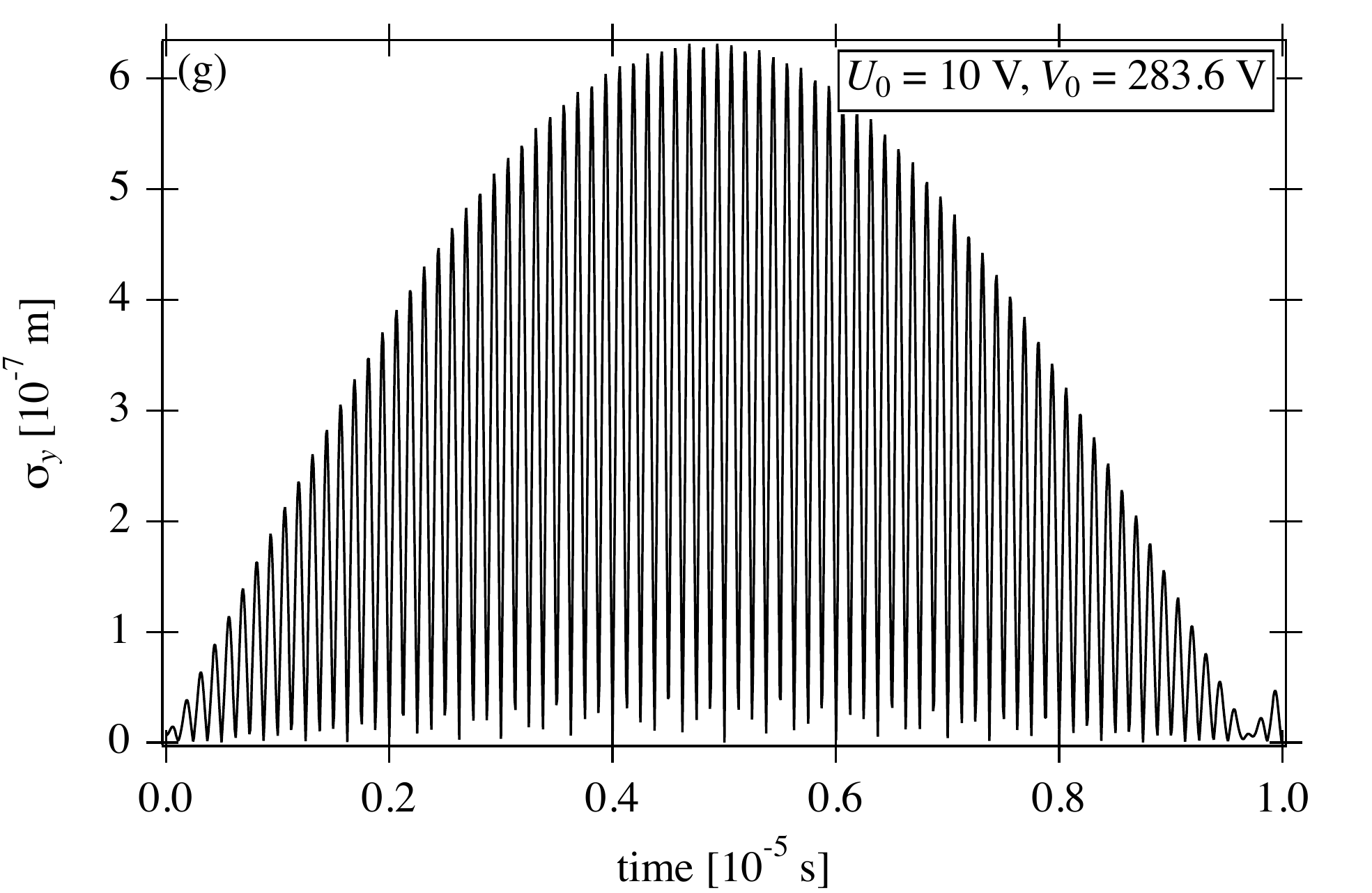}
\includegraphics[width=0.49\textwidth]{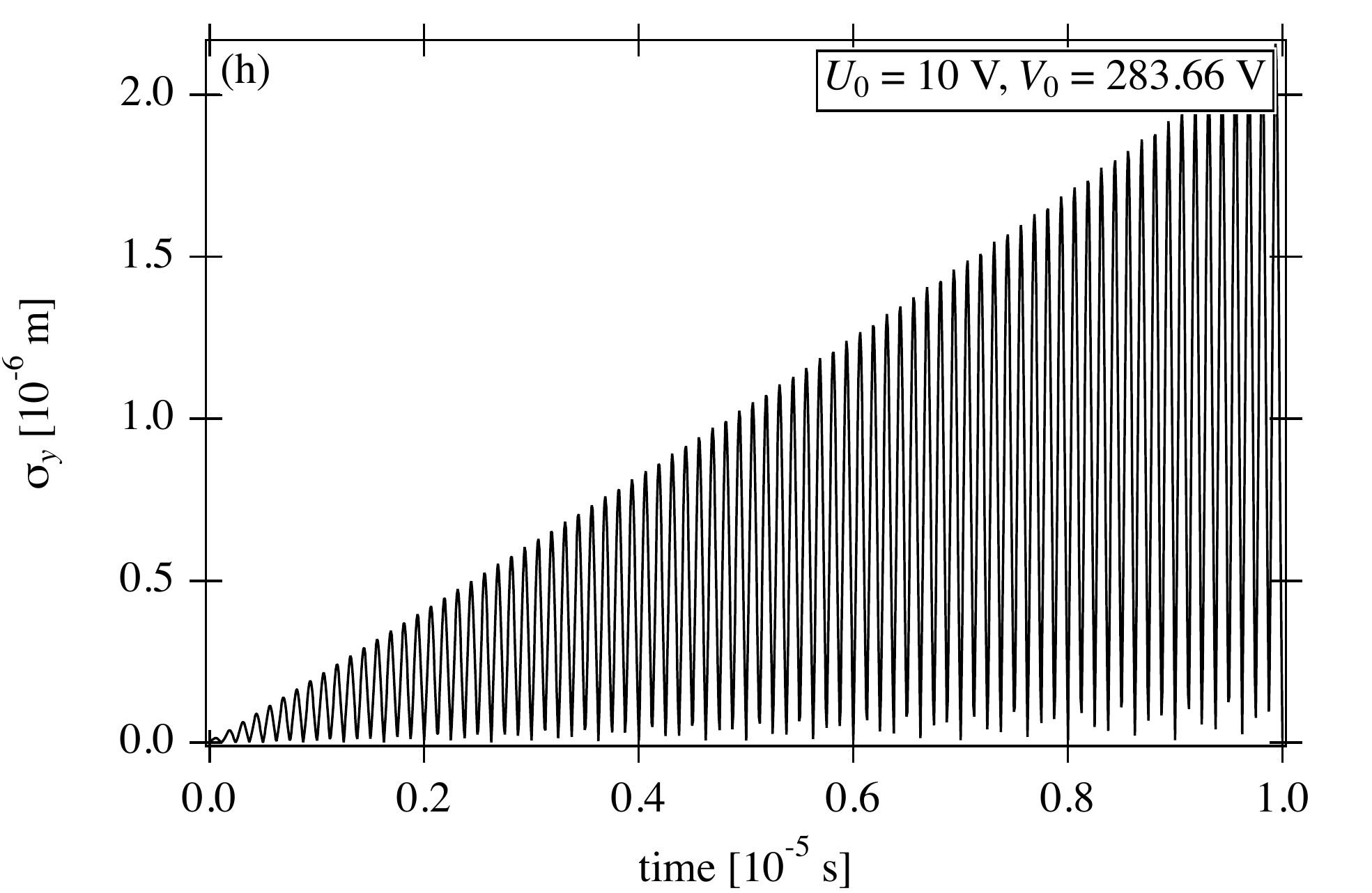}
\caption{\label{fig:283.659}Time evolution of the center-of-mass
  motion and width of the wave packet: (a), (c), (e), and (g) $U_0 =
  \SI{10}{V}$, $V_0 = \SI{283.6}{V}$ (inside the stability region);
  (b), (d), (f), and (h) $U_0 = \SI{10}{V}$, $V_0 = \SI{283.66}{V}$
  (outside the stability region).}
\end{figure}

\begin{figure}
  \begin{center}
    \includegraphics[width=0.65\textwidth]{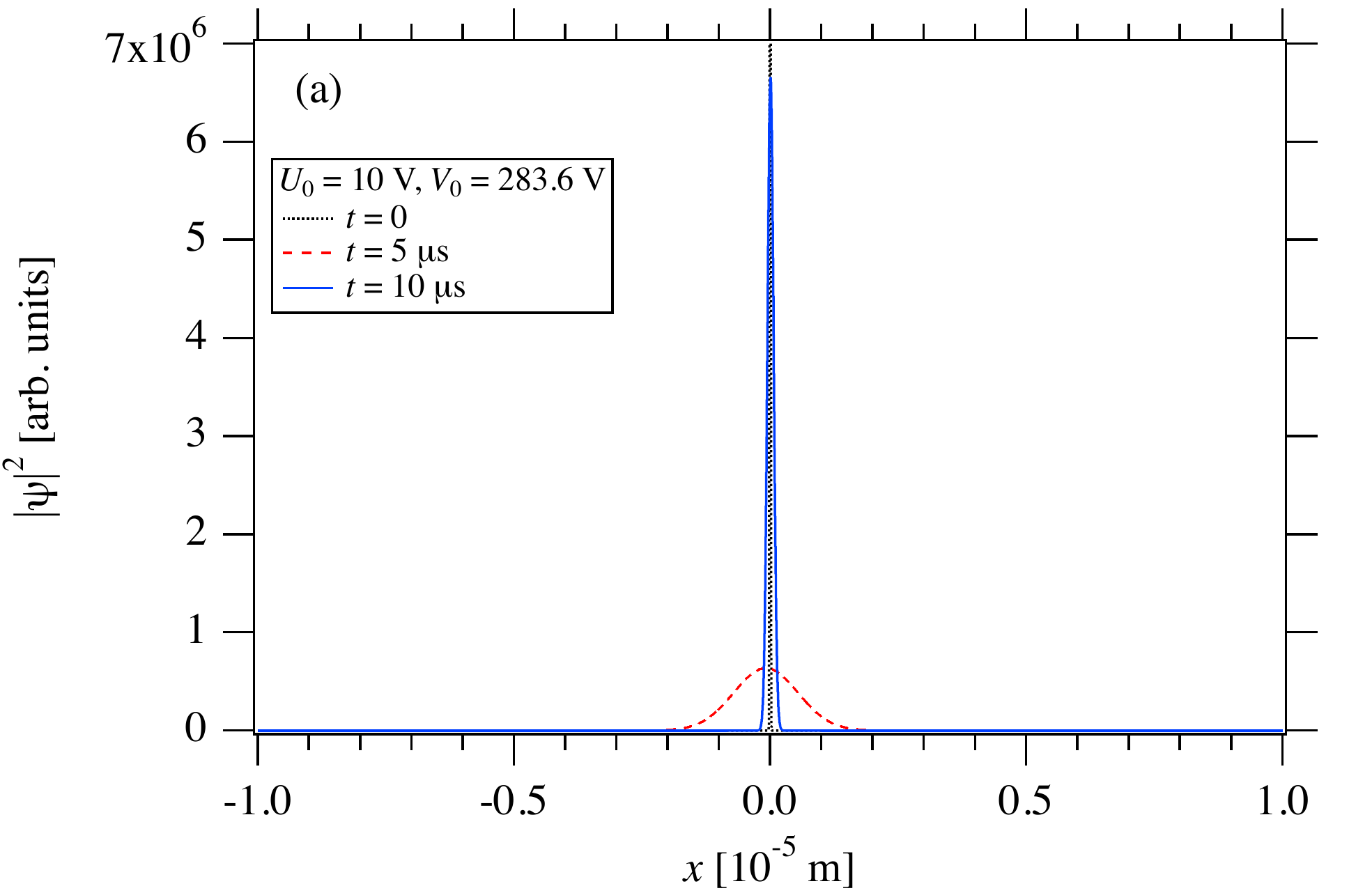}
    \includegraphics[width=0.65\textwidth]{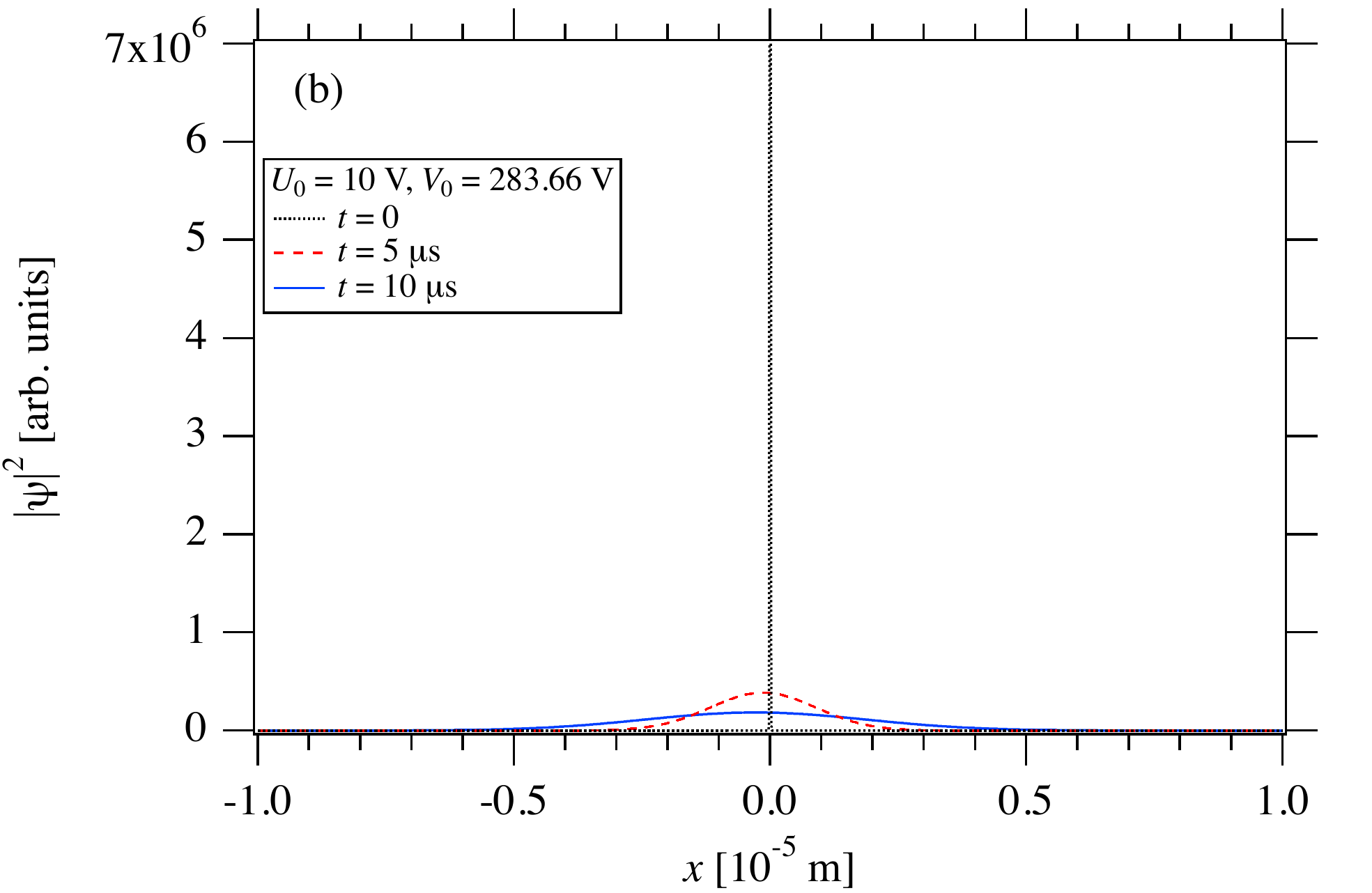}
  \end{center}
  \caption{\label{fig:snapshots}(Color online) Density distribution
    along the $x$ axis at times $t = 0$, $5$, and
    $\SI{10}{\micro\second}$ for: (a) the stable case $U_0 =
    \SI{10}{V}$, $V_0 = \SI{283.6}{V}$; (b) the unstable case $U_0 =
    \SI{10}{V}$, $V_0 = \SI{283.66}{V}$.  Note that, in both cases,
    the wave packet at $t=0$ peaks at $\left| \psi_x(x = 0) \right|^2
    \approx \num{5.7e7}$, beyond the scale of the figures.}
\end{figure}

We have checked this numerically by considering the border between
stable and unstable trajectories, see Fig.~\ref{fig:stability}.  Using
Mathematica,\cite{Mathematica_10} we find that the Mathieu
characteristic value $b_1$ for $U_0 = \SI{10}{V}$ is at $V_0 =
\SI{283.659}{V}$.  We ran simulations for points on either side of this
border, choosing $V_0 = \SI{283.6}{V}$ (stable) and $V_0 =
\SI{283.66}{V}$ (unstable), using now 1048576 grid points, in the
range $[\SI{-10}{\micro\meter}, \SI{10}{\micro\meter}]$, to account
for the wider oscillations of the wave packet.  The results of the
time evolution are
presented in Fig.~\ref{fig:283.659},
with snapshots of the density distribution given in
Fig.~\ref{fig:snapshots}.
We find that the trajectory is indeed bounded for $V_0 =
\SI{283.6}{V}$, along all dimensions (results for the z-axis are not
shown, as the motion is always simply harmonic along that direction).
The width of the wave packet, Figs.~\ref{fig:283.659}(e) and (g),
increases such that it extends farther than the center-of-mass
trajectory, but still presents a bounded oscillatory behavior.  In
contrast, for $V_0 = \SI{283.66}{V}$, the center-of-mass motion is
an unbounded oscillation, and the ion will eventually escape the
trapping region.  Of particular interest is the fact that the width of
the wave packet now also appears as an oscillation of increasing
amplitude.  This is particularly striking for the motion along the
y-axis, where the trajectory diverges much more slowly than along the
x-axis [due to the difference in the phase of the trapping field along
those two directions, see Eq.~(\ref{eq:rf-pot})], the width of the
wave packet grows with the same rate along both directions.

\subsubsection{Floquet approximate solution}

\begin{figure}
  \begin{center}
    \includegraphics[width=0.65\textwidth]{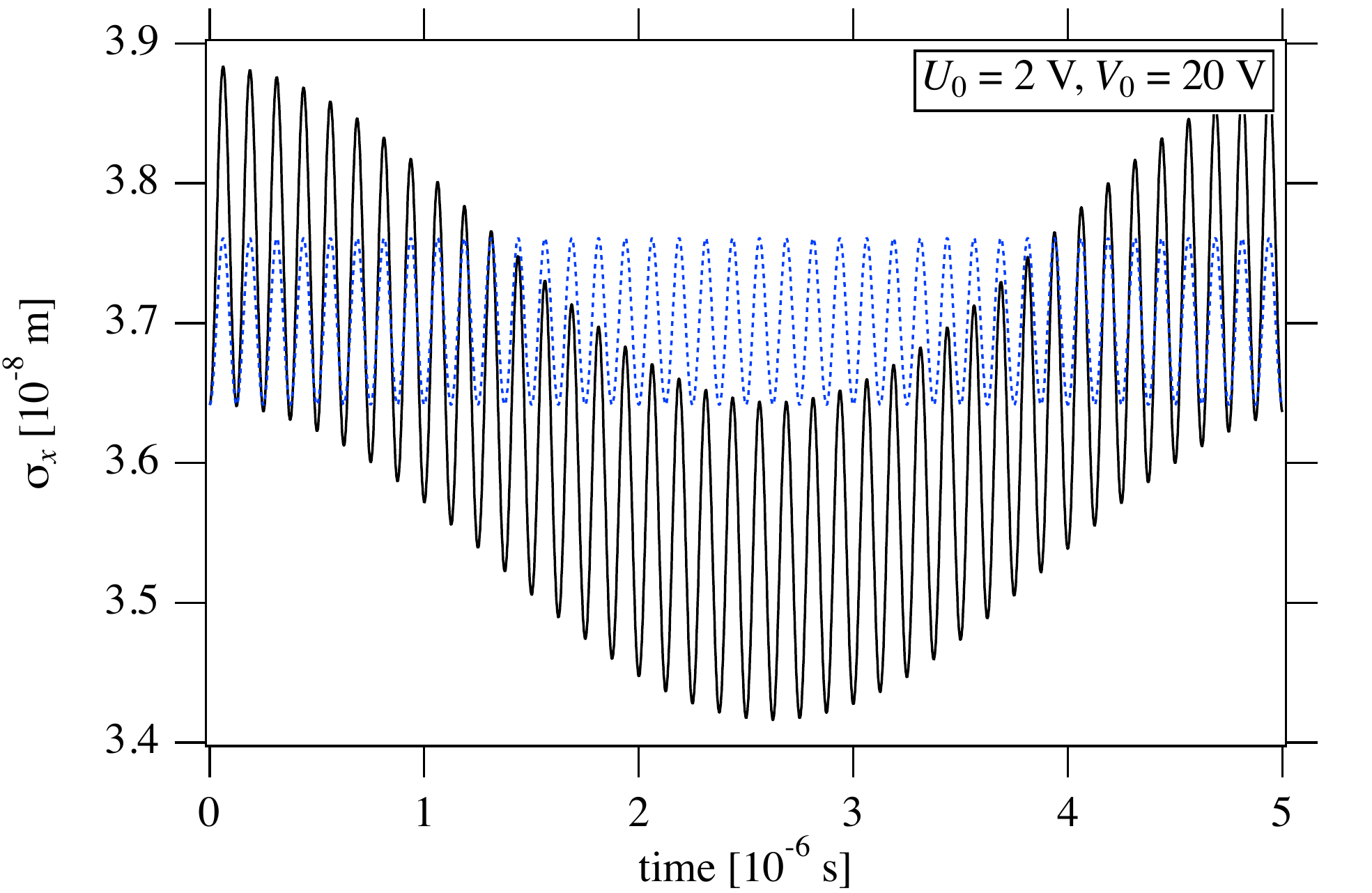}
  \end{center}
  \caption{\label{fig:leibfried}(Color online) Width $\sigma_x$ of the
    wave packet for $U_{0} = \SI{2}{V}$, $V_{0}
    = \SI{20}{V}$, for the approximate Floquet analytical solution
    Eq.~(\ref{eq:floquet_width}) (dashed line), compared to the
    corresponding value from the numerical solution of the full
    time-dependent Schr\"{o}dinger equation (solid line).}
\end{figure}

In Sec.~\ref{sec:floquet}, we presented the results of an approximate
analytical solution, based on the Floquet theorem, from which we
derived the a time-dependent width Eq.~(\ref{eq:floquet_width}).  We
will now compare the result thus obtained with the equivalent quantum
mechanical simulation.  In other to satisfy the conditions of the
approximate solution, namely $|a_x|, q_x^2 \ll 1$, we take $U_0 =
\SI{2}{\volt}$, $V_0 = \SI{20}{\volt}$, which corresponds to $a_x =
-0.00151$, $q_x = -0.0645$.  We see, Fig.~\ref{fig:leibfried},
that the approximate solution reproduces well the oscillation at the
trap frequency $\Omega$, but is off by a factor of $\sim 2$ in the
amplitude of the oscillation of the width.  It also doesn't reproduce
a longer-period oscillation that is present in the actual wave
packet. 

\subsection{Validity of the effective potential approximation}

\begin{figure}
  \begin{center}
    \includegraphics[width=0.65\textwidth]{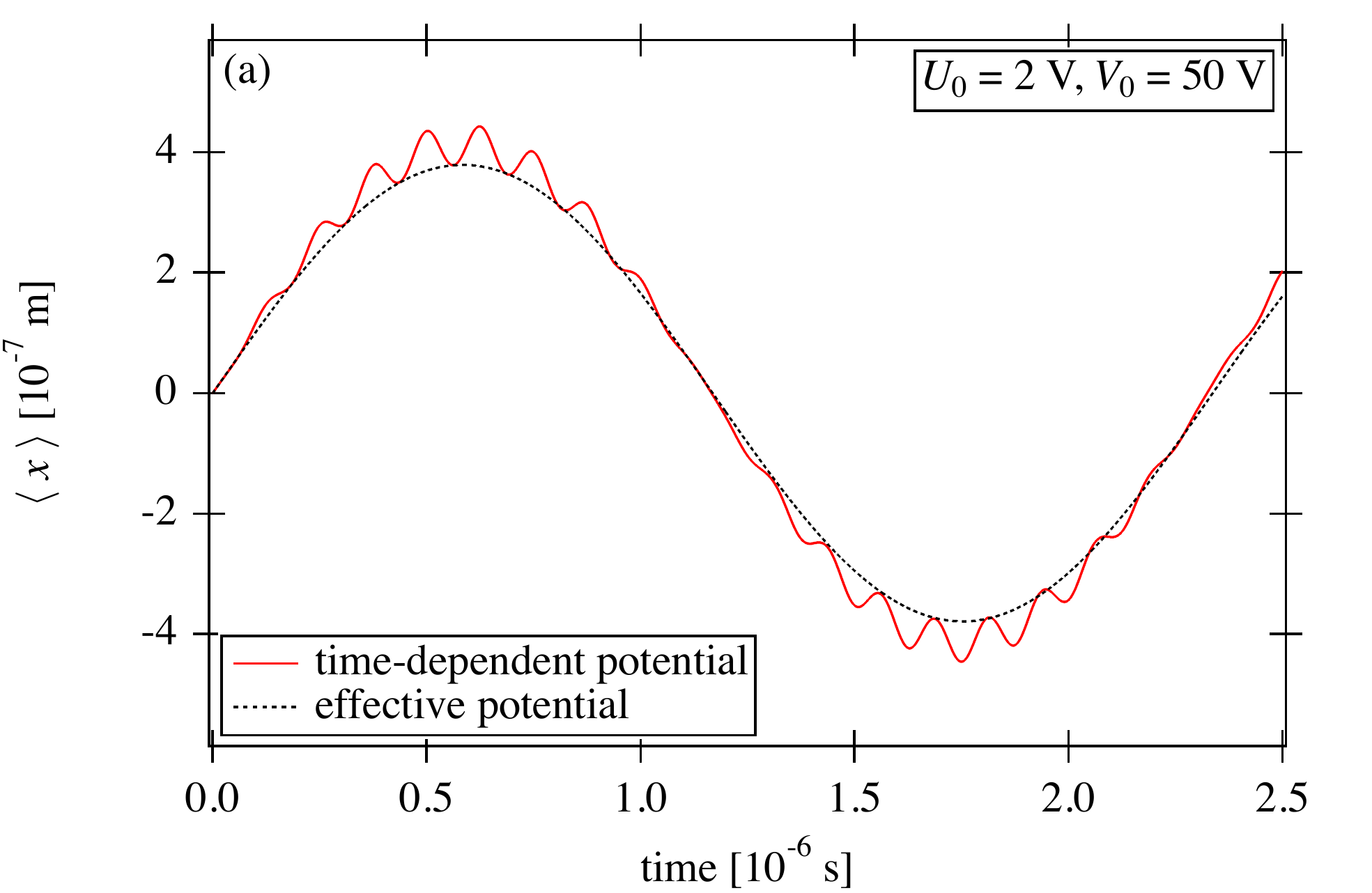}
    \includegraphics[width=0.65\textwidth]{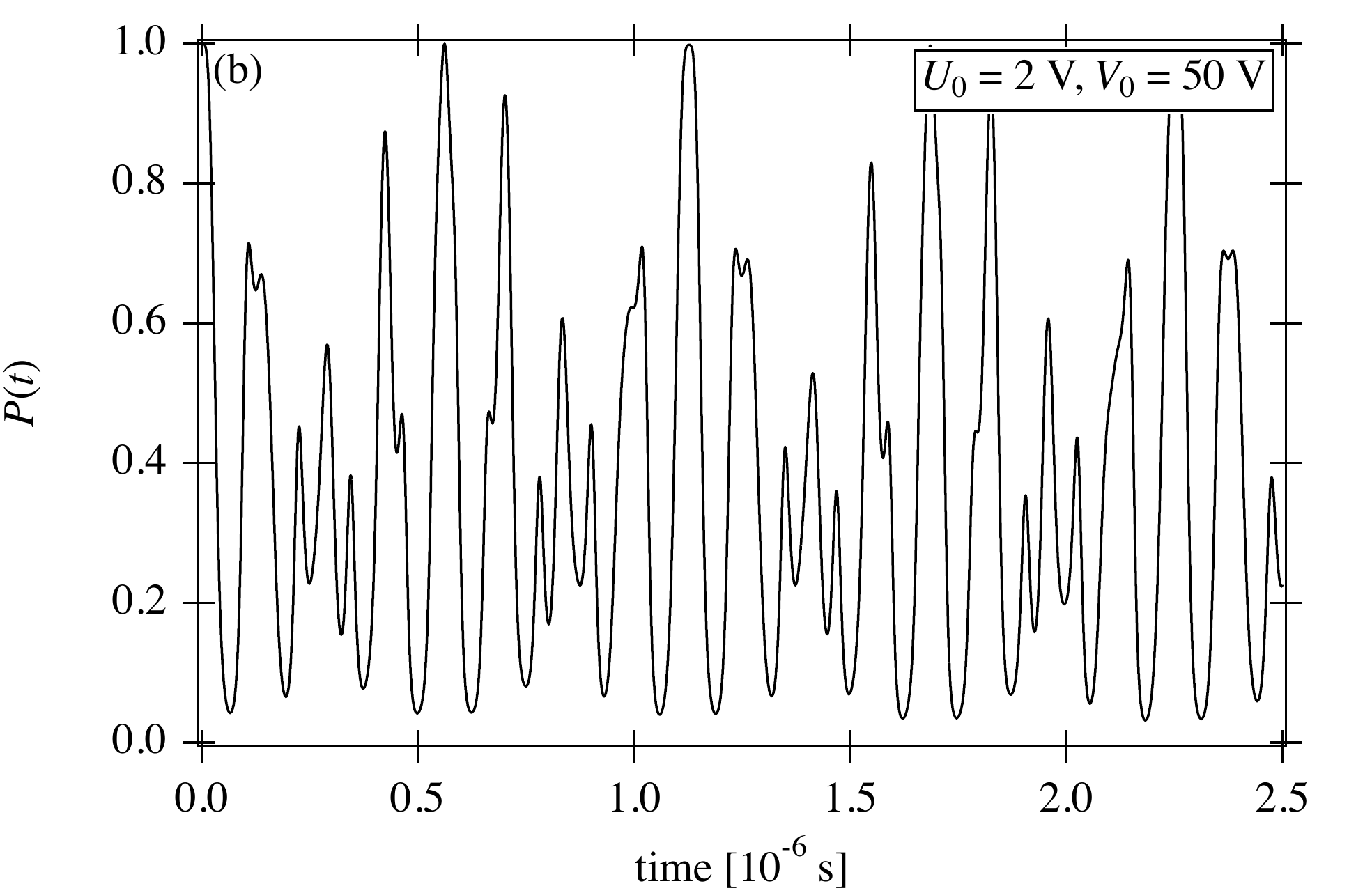}
  \end{center}
  \caption{\label{fig:proj_2_50}(Color online) (a) Time evolution of
    the center-of-mass motion calculated using the effective potential
    approximation vs.\ full quantum dynamics.  (b) Projection of the
    phase-corrected effective wave function $\psi_{\mathrm{eff},x}$ on the
    exact wave function.  The trapping field is $U_0 = \SI{2}{V}, V_0
    = \SI{50}{V}$ ($a \approx -1.5 \times 10^{-3}$, $q \approx
    0.45$).}
\end{figure}

For completeness, we also examine the the validity of effective
potential approximation, by calculating the effective wave functions
corresponding to $\psi_{\mathrm{eff}}$ [see Eq.~(\ref{eq:si_eff})] for
the same sets of $(U_{0}, V_{0})$ as in Sec.~\ref{sec:QvsC} before and
compared them with the actual time-dependent wave function by
calculating the projection
\begin{equation}
  \label{eq:proj}
    P(t) = \left| \Braket{ \psi_x(x,t) |\psi_{\mathrm{eff},x}(x,t)}
    \right|^{2}.
\end{equation}
Figure~\ref{fig:proj_2_50} presents results for $U_0 = \SI{2}{V}, V_0
= \SI{50}{V}$, which corresponds to $a \approx -1.5 \times 10^{-3}$,
$q \approx 0.45$; as mentioned in Sec.~\ref{sec:effective}, this
effective potential approximation is valid in the limit of small
absolute values of $a$ and $q$.\cite{Major_book_2005}
As expected, the effective potential solution reproduces the secular
motion of the ion, see Fig.~\ref{fig:proj_2_50}(a), but not the
micromotion.  The exact solution is recovered every quarter period of
the oscillation, as the projection $P(t)$ shows
[Fig.~\ref{fig:proj_2_50}(b)], but in between the discrepancy between
the exact and the effective solution can be almost complete.  The
non-corrected effective wave function $\phi_{\mathrm{eff}}$ is worse
on average, although it does present more singular times where the
projection $P(t) \approx 1$.

\begin{figure}
  \begin{center}
    \includegraphics[width=0.65\textwidth]{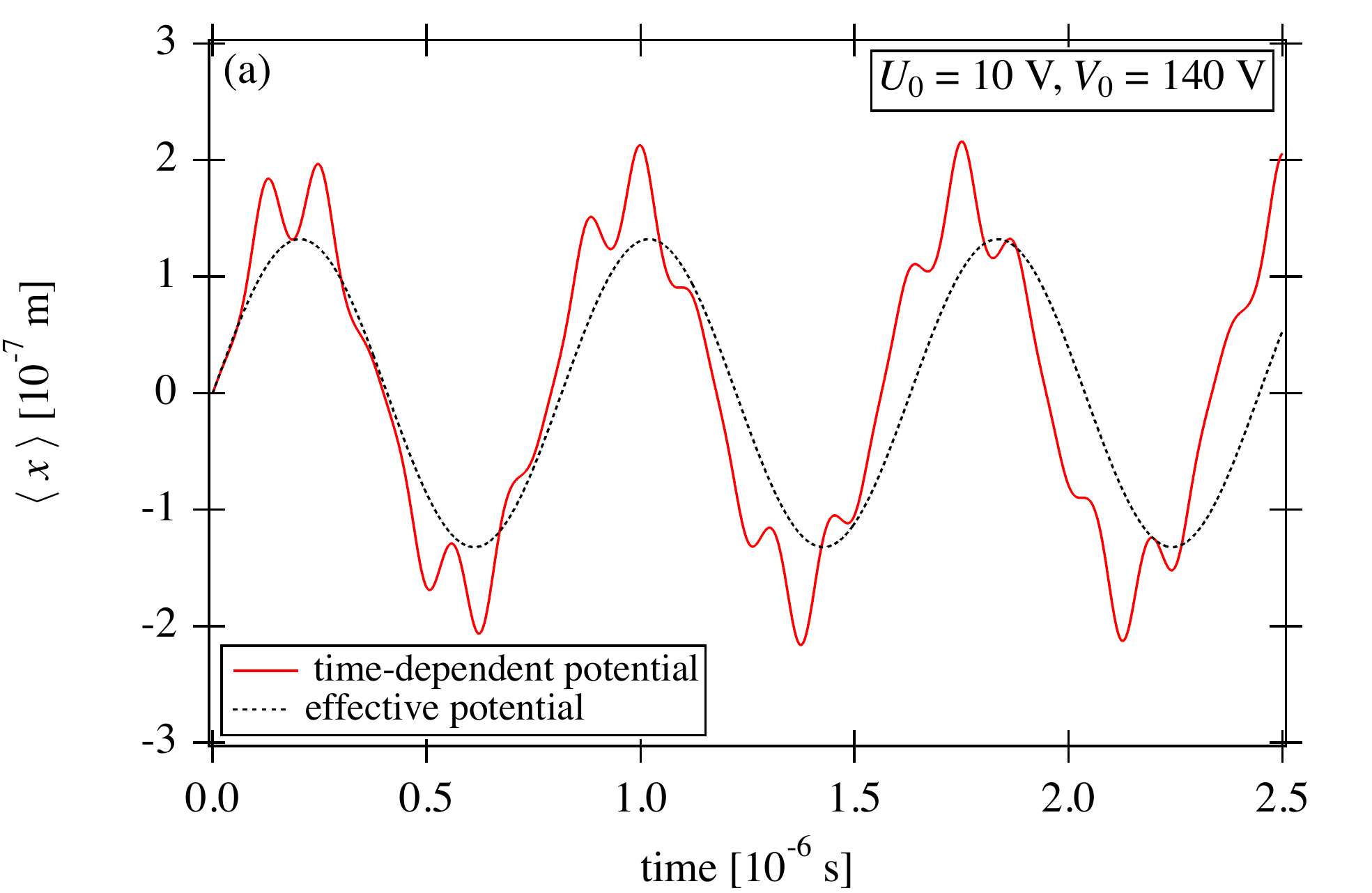}
    \includegraphics[width=0.65\textwidth]{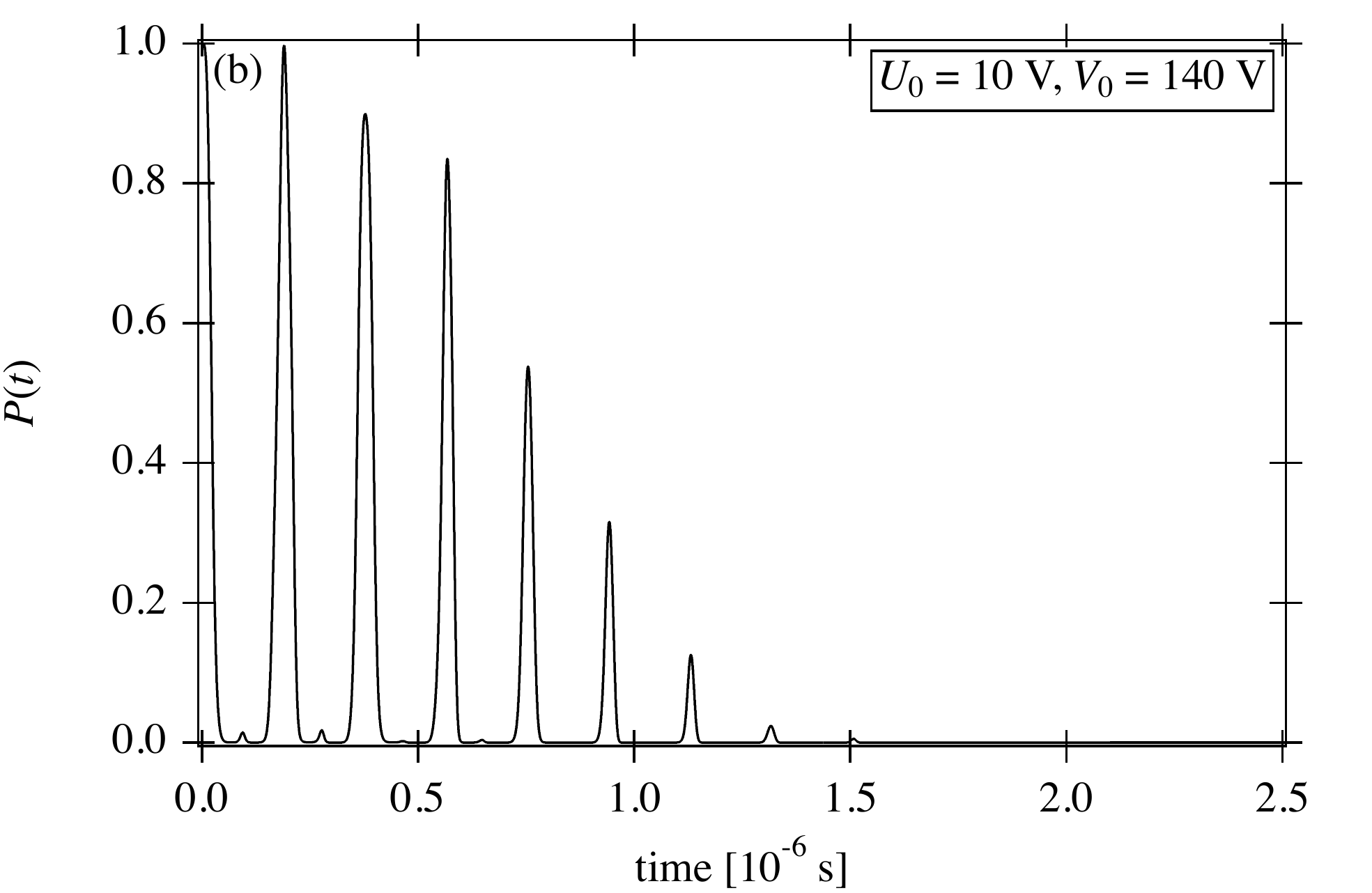}
  \end{center}
  \caption{\label{fig:proj_10_140}(Color online) (a) Time evolution of
    the center-of-mass motion calculated using the effective potential
    approximation vs.\ full quantum dynamics.  (b) Projection of the
    phase-corrected effective wave function $\psi_\mathrm{eff}$ on the
    exact wave function.  The trapping field is $U_0 = \SI{10}{V}, V_0
    = \SI{140}{V}$ ($a \approx -7.6 \times 10^{-3}$, $q \approx
    0.16$).}
\end{figure}

When increasing the trapping field to $U_0 = \SI{10}{V}, V_0 =
\SI{140}{V}$, for which $a \approx -7.6 \times 10^{-3}$ and $q \approx
0.16$, the effective potential approximation is no longer applicable,
as can be seen in Fig.~\ref{fig:proj_10_140}.
The amplitude of the micromotion is so important that the trajectories
appear to different oscillation periods (a similar effect can be seen
in the width of the packet\cite{Li_PRA_1993}).  After a short time,
$\psi_\mathrm{eff}$ differs completely from the exact wave function,
as seen in Fig.~\ref{fig:proj_10_140}(b), although a revival in $P(t)$
is expected when the secular motion and the micromotion resynchronize.

\section{Conclusion}
\label{sec:conclusion}

We have performed quantum simulations of the full dynamics of an ion
in a linear Paul trap, including the time-dependent variation of the
trapping potential.  We have shown that the center-of-mass motion of
the ion is well reproduced by the classical equations of motion, even
outside the so-called stability region, where the trajectories are no
longer bounded.\cite{Major_book_2005}  The use of the effective
potential approximation\cite{Major_book_2005,Cook_PRA_1985} allows
one to get the average (secular) motion of the center of mass correct,
but completely neglects the micromotion, which can be on par with the
secular motion for certain trapping parameters, to a point where the
results are completely different from the exact dynamics
(Fig.~\ref{fig:proj_10_140}).

Considering the oscillations of the width wave packet, we have found
that they are bounded for stable trajectories, even when the wave
packet is not Gaussian.  This extends previous work that had
demonstrated oscillations only for coherent wave
packets,\cite{Li_PRA_1993,Leibfried_RMP_2003,Glauber_1992,Glauber_1992b}
and will merit further investigation.  The wave packet was also seen
to continuously increase in width for conditions outside the stability
region, while it remained bounded for stable trajectories.  Our
results confirm the Floquet analysis of the quantum dynamics that
shows that the center of mass of the wave packet should follow the
classical trajectory,\cite{Combescure_AIHPA_1986,Brown_PRL_1991} and
that the stability criterion based on classical trajectories also
applies to quantum motion.\cite{Li_PRA_1993,Combescure_AIHPA_1986}

This study opens up the possibility of using semi-classical models for
the simulation of trapped ions.  The center-of-mass motion could be
considered to be classical, while internal degrees of freedom could be
treated quantum mechanically.  This could greatly simplify, for
instance, an extensive treatment of the interaction of a trapped atomic ion
with laser pulses, or simulations of the trapping of molecular ions.

\section*{Acknowledgments}
The simulations were performed on resources provided by the Swedish
National Infrastructure for Computing (SNIC) at the High Performance
Computing Center North (HPC2N).  Financial support from Ume{\aa}
University is gratefully acknowledged.


\end{document}